\documentclass[letterpaper, 12pt]{article}

\usepackage[dvipsnames]{xcolor}

\usepackage{
  amsmath, amsthm, amssymb, mathtools, dsfont, units,          
  graphicx, wrapfig, subfig, float,                            
  listings, color, inconsolata, pythonhighlight,               
  fancyhdr, sectsty, hyperref, enumerate, enumitem, framed }   

\usepackage{hyperref} 

\usepackage{newpxtext, newpxmath, inconsolata}
\usepackage{algorithm}
\usepackage{algpseudocode}
\usepackage{tikz}
\usepackage{tikz-network}
\usepackage{enumitem}
\usepackage[title,titletoc]{appendix}

\usepackage[left=1.35in, right=1.35in, top=1.0in, bottom=.9in, headsep=.2in, footskip=0.35in]{geometry}

\usepackage[bottom]{footmisc}

\allowdisplaybreaks

\usepackage[font={it,footnotesize}]{caption}

\hypersetup{colorlinks=true, linkcolor=RoyalBlue, citecolor=RedOrange, urlcolor=ForestGreen}

\usepackage{titlesec}
\titleformat{\section}{\large\bfseries}{\thesection}{1em}{}

\setlist[itemize]{wide=0pt, leftmargin=16pt, labelwidth=10pt, align=left}

\usepackage[subfigure]{tocloft}

\cftsetindents{section}{0em}{0em}

\makeatletter
\renewcommand{\cftsecpresnum}{\begin{lrbox}{\@tempboxa}}
\renewcommand{\cftsecaftersnum}{\end{lrbox}}
\makeatother

\graphicspath{{Images/}{../Images/}}

\usepackage[mathscr]{euscript}

\theoremstyle{definition}
\newtheorem{theorem}{Theorem}[section]

\newtheorem{assumption}[theorem]{Assumption}

\newtheorem*{theorem*}{Theorem}
\newtheorem*{proposition*}{Proposition}
\newtheorem*{lemma*}{Lemma}
\newtheorem*{corollary*}{Corollary}
\newtheorem*{definition*}{Definition}
\newtheorem*{example*}{Example}
\newtheorem*{remark*}{Remark}

\begin{document}

\def\smfbyname{}

\title{Markov-Functional Models with Local Drift}

\author{\stepcounter{footnote}\normalsize ShengQuan Zhou\thanks{Quantitative Research, Bloomberg L.P., New York, NY. Email: szhou80@bloomberg.net}}

\date{\normalsize \today}

\maketitle

\begin{abstract}
We introduce a Markov-functional approach to construct local volatility models that are calibrated to a discrete set of marginal distributions. The method is inspired by and extends the volatility interpolation of Bass (1983) and Conze and Henry-Labord\`{e}re (2022). The method is illustrated with efficient numerical algorithms in the cases where the constructed local volatility functions are: (1) time-homogeneous between or (2) continuous across, the successive maturities. The step-wise time-homogeneous construction produces a parsimonious representation of the local volatility term structure.
\end{abstract}


\section{Introduction}

The interpolation and extrapolation of a two-dimensional implied volatility surface is known to be a non-trivial problem. For a partial list of desirable properties one wishes to preserve in the construction of implied volatility surface, see Carr (2010) \cite{carr2010}. The solutions fall into two broad categories: the best-fit approach by a parametric model (e.g., Heston, SABR) and the non-parametric approaches. For the non-parametric approaches, previous attempts are made in two ways: (1) a two-step approach of fitting of an implied volatility skew in strike dimension followed by an interpolation in the maturity dimension \cite{carr2010, andreasen2011, lipton2011, phl2019}; (2) a full-scale global approach where a non-parametric representation is fitted directly to all observed option prices \cite{avellaneda1997,verma1999}. Dupire (1994) \cite{dupire1994}, Derman and Kani (1994) \cite{derman1994} in a discrete-time setting, show that it is possible to construct a \textit{local volatility model} that is consistent with a full surface of European option prices as a function of maturity and strike, while, in practice, only a discrete set of which is observed. In this paper, we consider the problem of volatility interpolation in the maturity dimension, given a discrete set of one-dimensional marginals. Our approach is inspired by and extends the construction of local volatility model of Bass (1983) \cite{bass1983} and Conze and Henry-Labord\`{e}re (2022) \cite{phl2022}, calibrated exactly to one or more marginals. The step-wise time-homogeneous construction described in Section \ref{krylov_section} produces a parsimonious representation of the local volatilities. 

Mimicking certain features of an It\^{o} process by the solution of a stochastic differential equation (SDE) is a recurring topic of interest. Krylov (1985) \cite{krylov1985} constructs a time-homogeneous diffusion that possesses the same Green $\lambda$-measure as a general It\^{o} process, where a Green $\lambda$-measure of a process records the average amount of time that the process spends in each Borel set before being killed at the first jump of an independent Poisson process with intensity $\lambda$. Gy\"{o}ngy (1986) \cite{gyongy1986} proves that there exists a time-dependent diffusion process with non-random coefficients that matches the one-dimensional marginal distributions of an It\^{o} process. Brunick and Shreve (2013) \cite{shreve2013} extend Gy\"{o}ngy (1986) by showing that the mimicking process can preserve not only the marginal distributions, but also the joint distribution of certain functionals of the It\^{o} process (e.g., running maximum, running average) at each fixed time. More recently, Lacker, Shkolnikov, and Zhang (2019) \cite{lacker2019} inverts the Markovian projection of Gy\"{o}ngy (1986) \cite{gyongy1986} by finding a different It\^{o} process with the fixed-time marginal distributions matching those of a given one-dimensional diffusion. Another line of work is devoted to a similar problem: mimicking an It\^{o} process that is a sub-martingale by a Markov process, not restricted to the solutions of SDE. Kellerer (1972) \cite{kellerer1972} establishes the existence of the Markov solutions matching pre-specified marginals, extended to multiple dimensions by Pammer, Robinson, and Schachermayer (2023) \cite{pammer2023}. Madan and Yor (2002) \cite{madan2002} provide explicit constructions of such Markov processes using three methods: Skorokhod embedding, time-changed Brownian motion by inhomogeneous independent increments, and continuous construction by Dupire (1994) \cite{dupire1994}.

The paper is organized as follows. In Section \ref{method}, we introduce the framework of Markov-functional models with local drift, its specializations and fixed-point numerical algorithms. In Section \ref{example}, we illustrate the methods by two sets of examples: one case with synthetic double exponential distributions and another with market data. In Section \ref{conclusion}, we summarize and conclude. In Appendix \ref{skew_fitting}, we describe the method we used to fit the implied volatility skews, hence the marginal distributions, slice by slice such that no-arbitrage conditions are enforced in both the strike dimension and the maturity dimension. 

\section{Markov-Functional Models by Local Drift}
\label{method}

Consider a horizon $T$ and a discrete set of marginal distributions $\nu_i$ on $\mathbb{R}$, $1\le i\le n$, implied in practice from market prices of vanilla options with maturities $0\triangleq T_0 < T_1 < \cdots < T_n \triangleq T$ and free of calendar arbitrage. To construct a (real-valued) martingale price process $(S_t)_{0\le t\le T}$, that meets the given marginal distributions, we assume that $S_t$ can be written as a Markov functional \cite{bergomibook}:
\begin{align}
\label{flowfunction}
S_t = f(t,X_t),
\end{align}
where $(X_t)_{0\le t\le T}$ is a process driven by a Brownian motion $(W_t)_{0\le t\le T}$ and guided by a \textit{local drift function} $\mu(t,x)$:
\begin{align}
dX_t = \mu(t,X_t)dt + dW_t, \quad X_0\in \mathbb{R}.
\end{align}
The forward density $p(t,x)$ of $X_t$ is known to solve the Fokker-Planck equation:
\begin{align}
\label{ppde}
\left[\frac{\partial}{\partial t} + \frac{\partial}{\partial x}\mu - \frac{1}{2}\frac{\partial^2}{\partial x^2}\right]p(t,x) = 0.
\end{align}
Applying It\^{o}'s formula to $f(t,X_t)$ yields:
\begin{align}
\label{itolemma}
    df(t,X_t) = \left(\frac{\partial f}{\partial t} + \mu \frac{\partial f}{\partial x}+ \frac{1}{2}\frac{\partial^2 f}{\partial x^2}\right)dt + \frac{\partial f}{\partial x} dW_t.
\end{align}
We gather from Eq.(\ref{itolemma}) and the martingale property of $S_t$ that $f(t,x)$ satisfies:
\begin{align}
\label{fpde}
\left[\frac{\partial}{\partial t} + \mu \frac{\partial}{\partial x} + \frac{1}{2}\frac{\partial^2}{\partial x^2}\right]f(t,x) = 0.
\end{align}
To borrow a terminology from probabilistic modeling and inference, we call  the process $(X_t)_{0\le t\le T}$ the \textit{flow variable} and the function $f(t,x)$ the \textit{flow function}. Correspondingly, the process $S_t$ is called the \textit{target variable} and $\nu_1,\cdots,\nu_n$ the \textit{target distributions}. Given a drift function $\mu(t,x)$ and the initial conditions
\begin{align}
p(0,x) = \delta(x-X_0), \quad f(0,X_0) \equiv S_0,
\end{align}
where $\delta(\cdot)$ is the Dirac-$\delta$ function, one can solve Eq.(\ref{ppde}) and Eq.(\ref{fpde}) under appropriate boundary conditions or terminal conditions. Specifically, to meet the given marginal distributions $S_{T_i}\sim \nu_i$ by quantile-matching:
\begin{align}
\label{quantilematching}
f(T_i, x) = F_{\nu_i}^{-1}\circ F_{X_{T_i}}(x), \quad i=1,\cdots,n,
\end{align}
where $F_{\nu_i}(\cdot)$ is the cumulative distribution function (CDF) of the target variable $S_{T_i}$ and $F_{X_{T_i}}$ that of the flow variable $X_{T_i}$: $F_{X_{T_i}}(x) \triangleq \int_{-\infty}^x p(T_i,y)dy$. At $T_0=0$, the flow function $f(0,x)$ typically depends on the specific form of the model, e.g., the drift function $\mu(t,x)$. The resulting dynamics of $S_t=f(t,X_t)$ is a local volatility model:
\begin{align}
\label{localvolandflow}
dS_t = \partial_x f(t,f^{-1}(t,S_t))dW_t,
\end{align}
where $\partial_x f (t,x)\triangleq \frac{\partial f}{\partial x}(t,x)$. According to Eq.(\ref{quantilematching}), the flow function at a marginal maturity is monotonically increasing, i.e., $\partial_x f(T_i,x)\ge 0$. In fact, any local volatility model specified by a local volatility function $\sigma_{\text{loc}}(t,\cdot)$ can be written in an equivalent form of a flow function $f(t,x)$ determined by
\begin{align}
\label{flowfunctionode}
\frac{\partial f(t,x)}{\partial x} = \sigma_{\text{loc}}(t,f(t,x)) \Rightarrow \int \frac{df}{\sigma_{\text{loc}}(t,f)} = x.
\end{align}
Comparing Eq.(\ref{flowfunctionode}) with the BBF formula of Berestycki, Busca, and Florent \cite{bbf2004} gives the following implicit equation for the Markov flow function $f(0,\cdot)$ at $t=0$:
\begin{align}
\label{bbfrelation}
\frac{f-S_0}{\sigma_{\text{imp}}(T=0,f)} = x-X_0,
\end{align}
if the short-term implied volatility skew $\sigma_{\text{imp}}(T=0,\cdot)$ is sufficiently regular. In general, Eq.(\ref{quantilematching}), a relation between the flow function and the flow variable's distribution function, will not be satisfied by an arbitrary drift function $\mu(t,x)$. Instead, the drift function $\mu(t,x)$ (hence the forward distributions of the flow variable) and the flow function $f(t,x)$ need to solved jointly. A simple example of the flow function known in closed-form is the Black-Scholes model, where $f(t,x) = S_0 \exp\left( \sigma x - \frac{1}{2}\sigma^2 t \right)$ and $(X_t)_{0\le t\le T}$ is a Brownian motion. 

\vspace{3mm}

{\setlength{\parindent}{0cm}
\underline{\textit{Remark}}. If $(X_t)_{0\le t\le T}$ and $f(t,x)$ is a solution to Eq.(\ref{flowfunction}), then with any deterministic function $C(t)$ of time $t$, $\tilde{X}_t \triangleq X_t + C(t)$ and $\tilde{f}(t,x) \triangleq f(t, x - C(t))$ is also a solution. This ambiguity can be removed by enforcing $\mathbb{E}[X_t]=0$, $\forall t\in[0,T]$.
}

\subsection{Bass (1983)}
\label{bass_section}

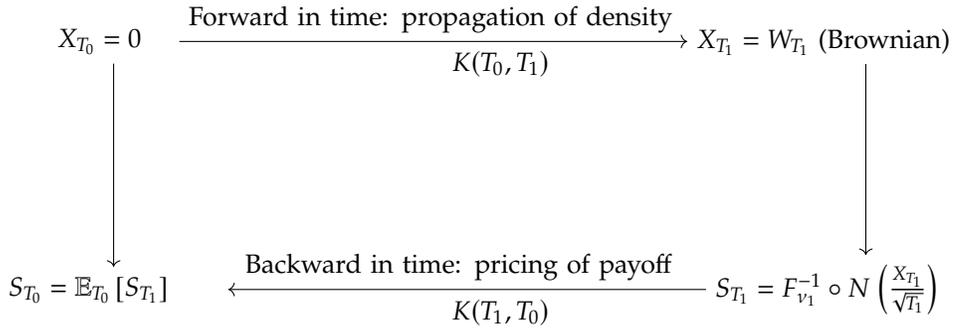
\begin{figure}[!ht]

\centering

{\footnotesize{
\begin{tikzpicture}

\node[text width=1.5cm] (p1) at (0,0) {$X_{T_0}=0$};
\node[text width=4.5cm] (p2) at (10,0) {$X_{T_1}=W_{T_1}$ (Brownian)};
\node[text width=4cm] (p3) at (10,-3.3) {$S_{T_1}=F_{\nu_1}^{-1}\circ N\left( \frac{X_{T_1}}{\sqrt{T_1}} \right)$};
\node[text width=2.8cm] (p4) at (0,-3.3) {$S_{T_0}=\mathbb{E}_{T_0}\left[ S_{T_1} \right]$};
\node[text width=8cm] at (5,0.3) {\footnotesize{Forward in time: propagation of density}};
\node[text width=1cm] at (5,-0.3) {\footnotesize{$K(T_0,T_1)$}};
\node[text width=6.5cm] at (5,-3.0) {\footnotesize{Backward in time: pricing of payoff}};
\node[text width=1cm] at (5,-3.6) {\footnotesize{$K(T_1,T_0)$}};

\begin{scope}[every path/.style={->}]
    \draw (p1) -- (p2);
    \draw (p2) -- (p3); 
    \draw (p3) -- (p4);
    \draw (p1) -- (p4);
\end{scope} 

\end{tikzpicture}
}}

\caption{A schematic derivation of the flow function for the first forward period $t\in[0, T_1)$.}
\label{first_period_chart}

\end{figure}

Bass (1983) \cite{bass1983} considers a single forward period $t\in[0,T_1)$, where the flow variable $X_t$ is a Brownian motion itself, with $X_0=0$. In other words, the drift function $\mu(t,x)\equiv 0$. In the first forward period $t\in[0,T_1)$, the distribution of the flow variable at time $t$ is simply a Gaussian $X_t\sim \mathcal{N}(0,t)$:
\begin{align}
F_{X_{t}}(x) = N\left(\frac{x}{\sqrt{t}}\right), \quad t\in[0,T_1),
\end{align}
where $N(\cdot)$ is the CDF of the standard normal distribution. By quantile-matching, the flow function at $t=T_1$ is determined:
\begin{align}
f(T_1,x) = F^{-1}_{\nu_1}\circ N\left(\frac{x}{\sqrt{T_1}}\right).
\end{align}
Because $\mu(t,x)\equiv 0$, the flow function satisfies a heat equation $\frac{\partial f}{\partial t} + \frac{1}{2}\frac{\partial^2 f}{\partial x^2} = 0$ for $t\in[0,T_1)$ and is a convolution of its terminal values at $T_1$ with the heat kernel propagator from $T_1$ back to $t$,
\begin{align}
\label{bass_quadrature_1}
f(t,x) = K(T_1,t)\star f(T_1,x) \triangleq \int_{-\infty}^{+\infty} K_{T_1-t}(y)f(T_1,x-y)dy,
\end{align}
where $K_\tau(x) \triangleq \frac{1}{\sqrt{2\pi \tau}}e^{-\frac{x^2}{2\tau}}$. The above procedure is illustrated in Figure (\ref{first_period_chart}).

Note that, in the continuous-time limit of an appropriate discretization of the target distribution in binomial quantiles, the implied binomial tree of Rubinstein (1994) \cite{rubinstein1994} corresponds to the Bass martingale $f(t,W_t)$, $t\in[0,T_1]$.

\subsection{Conze and Henry-Labord\`{e}re (2022)}
\label{phl_section}

The problem of constructing a Markov martingale that is closest to a Brownian motion among those with prescribed initial and terminal distributions was introduced and studied by Backhoff-Veraguas \textit{et al}. (2020) \cite{backhoff2020}. The solution is known as the \textit{standard stretched Brownian motion}. Conze and Henry-Labord\`{e}re (2022) \cite{phl2022} extends the construction of Bass (1983) \cite{bass1983} to the subsequent forward periods. For each forward period $t\in [T_i, T_{i+1})$, $i=1,2,\cdots,n-1$, the flow variable $X_t$ still has Brownian dynamics $dX_t = dW_t$, but in this case combined with an uncertain initial value $X_{T_i}$. The probability distribution of the initial value $X_{T_i}$ is determined jointly with the flow function for the period $t\in [T_i, T_{i+1})$ by the following fixed-point equation:
\begin{align}
\label{phl_quadrature_1}
F_{X_{T_i}} = F_{\nu_i}\circ K(T_{i+1},T_i) \star F_{\nu_{i+1}}^{-1}\circ K(T_i, T_{i+1})\star F_{X_{T_i}}.
\end{align}
The flow function for $t\in [T_i, T_{i+1})$ is obtained afterwards:
\begin{align}
\label{phl_quadrature_2}
f(t,x) =  K(T_{i+1},t) \star F_{\nu_{i+1}}^{-1}\circ K(T_i, T_{i+1})\star F_{X_{T_i}}(x).
\end{align}
The above procedure is illustrated in Figure (\ref{second_period_chart}) for $t\in[T_1,T_2)$:

\begin{figure}[H]

\centering

{\footnotesize{
\begin{tikzpicture}
\node[text width=1cm] (p1) at (0,0) {$X_{T_1}$};
\node[text width=5cm] (p2) at (10,0) {$X_{T_2}=X_{T_1} + W_{T_2} - W_{T_1}$};
\node[text width=4cm] (p3) at (10,-3.3) {$S_{T_2}=F_{\nu_2}^{-1}\circ F_{X_{T_2}}(X_{T_2})$};
\node[text width=5.5cm] (p4) at (0,-3.3) {$S_{T_1}=F_{\nu_1}^{-1}\circ F_{X_{T_1}}(X_{T_1})=\mathbb{E}_{T_1}\left[ S_{T_2} \right]$};
\node[text width=8cm] at (5,0.3) {\footnotesize{Forward: propagation of density}};
\node[text width=1cm] at (5,-0.3) {\footnotesize{$K(T_1,T_2)$}};
\node[text width=4.5cm] at (5.5,-3.0) {\footnotesize{Backward: pricing of payoff}};
\node[text width=1cm] at (5,-3.6) {\footnotesize{$K(T_2,T_1)$}};

\begin{scope}[every path/.style={->}]
    \draw (p1) -- (p2);
    \draw (p2) -- (p3); 
    \draw (p3) -- (p4);
    \draw (p1) -- (p4);
\end{scope} 
\end{tikzpicture}
}}
\caption{A schematic derivation of the flow function for the second forward period $t\in[T_1,T_2)$.}
\label{second_period_chart}
\end{figure}
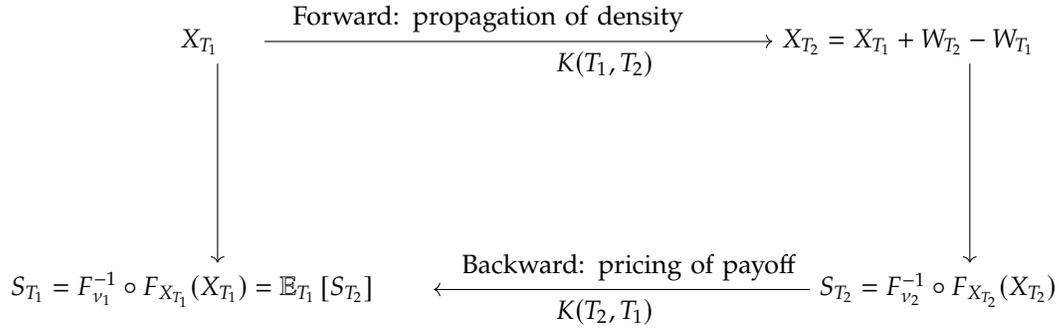

More recently, Acciaio \textit{et al}. (2023) \cite{acciaio2023} shows the existence and uniqueness of the solution to Eq.(\ref{phl_quadrature_1}) and prove that this model converges to the Dupire local volatility model when the mesh size $\sup_{i=1,\cdots,n}|T_i-T_{i-1}|$ converges to zero. Joseph \textit{et al}. (2023) provides a clear exposition on the link between the fixed-point iteration of Eq.(\ref{phl_quadrature_1}) and the measure-preserving martingale Sinkhorn's system. The flow variable in Bass (1983) \cite{bass1983} and Conze and Henry-Labord\`{e}re (2022) \cite{phl2022} has Brownian dynamics, which facilitates an efficient simulation of Monte Carlo paths. Moreover, the flow functions in the successive forward periods $[T_{i-1},T_{i})$, $i=1,2,\cdots,n$, can be solved in parallel. In particular, the distributions of the flow variable at the beginning of each forward period are determined separately. As a consequence, the overall flow function (hence the local volatility function) is, in general, not continuous across the marginal maturities. In a Monte Carlo simulation, the flow variable $X_{T_i}$ at the beginning of each forward period $[T_i,T_{i+1})$, $i=1,2,\cdots,n-1$, must be path-wisely determined by the continuity condition:
\begin{align}
\label{continuity1}
S_{T_i} = S_{T_{i}^-} \Rightarrow X_{T_i} = f^{-1}(T_i, f(T_i^-, X_{T_i^-})).
\end{align}
Finally, the flow function in this construction is explicitly time-dependent, which may lead to unrealistic term structures in the local volatility interpolation.

\subsection{Step-wise Time-Homogeneous Diffusion} 
\label{krylov_section}

We construct a step-wise time-homogeneous (i.e., autonomous) diffusion process, in the spirit of Krylov (1985) \cite{krylov1985}, that, instead of replicating the Green $\lambda$-measure, meets a discrete set of marginal distributions. In the sense of a generalized diffusion, the existence of a time-homogeneous local volatility model in $t\in[0,T]$ that meets a given marginal at $t=T$ is established by Noble (2013) \cite{noble2013}. Our approach is to start with a time-homogeneous flow function:
\begin{align}
S_t = f(X_t),
\end{align}
where the flow variable has a time-homogeneous drift $\mu(x)$:
\begin{align}
dX_t = \mu(X_t)dt + dW_t.
\end{align}
The dynamics of $S_t$ is a manifestly time-homogeneous local volatility model:
\begin{align}
dS_t = f'(f^{-1}(S_t))dW_t,
\end{align}
where $f'(x)\triangleq \frac{df(x)}{dx}$, $f''(x)\triangleq \frac{d^2f(x)}{dx^2}$. The martingale property of $S_t$ given in Eq.(\ref{fpde}) imposes the following relation between the flow function and the drift function:
\begin{align}
\label{krylovdrift}
\mu(x) = - \frac{f''(x)}{2f'(x)} = -\frac{1}{2}\frac{\partial }{\partial x}\log f'(x), \quad \text{if} \,\, f'(x)>0.
\end{align}
The drift function $\mu(x)$ set by Eq.(\ref{krylovdrift}) is undefined when $f'(x)=0$ or $\infty$, which occurs when the probability density of the flow variable at $x$, or the target variable at $f(x)$, respectively, is exactly zero. For Algorithm \ref{krylov_fixed_point_1} and \ref{krylov_fixed_point_2} to work properly, we impose the following condition on the target distributions:

\begin{assumption}
 In Algorithm \ref{krylov_fixed_point_1} and \ref{krylov_fixed_point_2}, the target marginal cumulative distribution functions $F_{\nu_i}(\cdot)$, $i=1,2,\cdots,n$, are strictly increasing on $[0,\infty)$ or $(-\infty,+\infty)$. 
\end{assumption}

For the first forward period $t\in[0,T_1)$, we solve the following fixed-point equation starting with an initial guess of zero drift $\mu(x)\equiv 0$ (Algorithm \ref{krylov_fixed_point_1}):
\begin{align*}
\mu \xrightarrow{\text{solve Eq.(\ref{ppde})}} F_{X_{T_1}}
\xrightarrow{\text{quantile matching}}  f = F_{\nu_1}^{-1}\circ F_{X_{T_1}}
\xrightarrow{\text{update drift}} \mu =-\frac{f''}{2f'}
\end{align*}

\begin{algorithm}[!ht]
\caption{Time-homogeneous flow fixed-point iteration for $t\in[0,T_{1})$}\label{krylov_fixed_point_1}
\begin{algorithmic}
\State Start with an initial guess of $\mu(x)$
\Repeat
\State Solve the forward equation (\ref{ppde}) to obtain $F_{X_{T_1}}(x)$.
\State Set the flow function by $f(x) = F_{\nu_1}^{-1}\circ F_{X_{T_1}}(x)$.
\State Update drift function $\mu(x) = -\frac{f''(x)}{2f'(x)}$.
\Until{$\mu(x)$ and $f(x)$ converge.}
\end{algorithmic}
\end{algorithm}

For each subsequent forward period $t\in[T_i,T_{i+1})$, $i=1,2,\cdots,n-1$, we construct a step-wise time-homogeneous flow function. Similar to Conze and Henry-Labord\`{e}re (2022) \cite{phl2022}, the flow variable $X_t$ has an uncertain initial value $X_{T_i}$ whose probability distribution needs to be determined jointly with the drift function $\mu(x)$ and the flow function $f(x)$ according to the condition that the flow function is time-homogeneous. At $t=T_i$,
\begin{align*}
S_{T_i} = f(X_{T_i}) = F_{\nu_i}^{-1}\circ F_{X_{T_i}}(X_{T_i}),
\end{align*}
while at $t=T_{i+1}$,
\begin{align*}
S_{T_{i+1}} = f(X_{T_{i+1}}) = F_{\nu_{i+1}}^{-1}\circ F_{X_{T_{i+1}}}(X_{T_{i+1}})
= F_{\nu_{i+1}}^{-1}\circ \hat{K}^{\text{\tiny{forward}},\mu}_{T_i\rightarrow T_{i+1}} \left[F_{X_{T_i}} (X_{T_{i+1}})\right],
\end{align*}
where the forward propagator $\hat{K}^{\text{\tiny{forward}}}_{T_i\rightarrow T_{i+1}}$ is a solver of Eq.(\ref{ppde}), determined by the drift function $\mu(x)$. Based on the fact that there is only one flow function $f(x)$ in each period, the fixed-point equation for $F_{X_{T_i}}(x)$ reads
\begin{align}
F_{X_{T_i}}(x) = F_{\nu_i}\circ F_{\nu_{i+1}}^{-1}\circ \hat{K}^{\text{\tiny{forward}},\mu}_{T_i\rightarrow T_{i+1}} \left[F_{X_{T_i}} (x)\right].
\end{align}
The following joint fixed-point equation produces such a solution when the iteration converges (Algorithm \ref{krylov_fixed_point_2}):

\begin{algorithm}[H]
\caption{Time-homogeneous flow fixed-point iteration for $t\in[T_i,T_{i+1})$}\label{krylov_fixed_point_2}
\begin{algorithmic}
\State Start with an initial guess of $\mu(x)$ and $F_{X_{T_i}}(x)$
\Repeat
\State Update the CDF of $X_{T_i}$ by $F_{X_{T_i}}(x) \leftarrow F_{\nu_i} \circ  F_{\nu_{i+1}}^{-1}\circ \hat{K}^{\text{\tiny{forward}},\mu}_{T_i\rightarrow T_{i+1}} \left[F_{X_{T_i}}(x) \right]$.
\State Set the flow function by $f(x) = F_{\nu_i}^{-1}\circ F_{X_{T_i}}(x)$.
\State Update drift function $\mu(x) = -\frac{f''(x)}{2f'(x)}$.
\Until{$\mu(x)$, $f(x)$ and $F_{X_{T_i}}(x)$ converge.}
\end{algorithmic}
\end{algorithm}

The construction of a Krylov-style local volatility model described in this section trades the same degrees of freedom in each marginal distribution with a time-homogeneous local volatility function in the corresponding forward period:
\begin{align}
dS_t = \sigma_i(S_t)dW_t, \quad t\in[T_{i-1},T_i),
\end{align}
where $i=1,2,\cdots,n$. It has the benefit of avoiding unrealistic shapes in the local volatility interpolation in the dimension of maturity, but still generates discontinuities across market maturities. In a Monte Carlo simulation, the flow variable $X_{T_i}$ at the beginning of each forward period $[T_i,T_{i+1})$, $i=1,2,\cdots,n-1$, must be determined path-wisely by the continuity condition:
\begin{align}
\label{continuity2}
S_{T_i} = S_{T_i^-} \Rightarrow X_{T_i} = f_{[T_i,T_{i+1})}^{-1}\circ f_{[T_{i-1},T_i)}(X_{T_i}^-),
\end{align}
where $f_{[T_i,T_{i+1})}(\cdot)$ denotes the time-homogeneous flow function restricted to the forward period  $[T_i,T_{i+1})$.



\subsection{Continuous Flow across Maturities} 
\label{continuous_section}

A specification of term structure underlies every volatility interpolation scheme considered previously. In Section \ref{bass_section} and \ref{phl_section}, the term structure is implicitly encoded in the heat kernel propagator $K_\tau(x)$. In Section \ref{krylov_section}, the term structure is explicitly specified to be step-wise time-homogeneous. Local volatility surfaces constructed period by period $[T_{i-1},T_i]$ to meet the marginal distributions at $T_i$, $i=1,2,\cdots,n$, are typically discontinuous across the marginal maturities. To enforce continuity, one solution is to explicitly specify a term structure of the flow function that is continuous in time and across maturities, by interpolating the snapshots $f(T_i,\cdot)$ at the marginal maturities $T_i$, $i=1,2,\cdots,n$. For example, assume that $f(t,x) = a_i + \frac{b_i}{\sqrt{t}}$, $\forall t\in [T_{i},T_{i+1}]$, for constant $a_i$ and $b_i$,
\begin{align}
\label{singularterm}
f(t,x) = \frac{\sqrt{\frac{T_i T_{i+1}}{t}}-\sqrt{T_i}}{\sqrt{T_{i+1}}-\sqrt{T_i}}f(T_i,x) + 
\frac{\sqrt{T_{i+1}}-\sqrt{\frac{T_i T_{i+1}}{t}}}{\sqrt{T_{i+1}}-\sqrt{T_i}}f(T_{i+1},x),
\end{align}
where the snapshots $f(T_i,x)$ are, as usual, determined by the quantile-matching condition in Eq.(\ref{quantilematching}). At $t=0$, the flow function is typically undetermined except the requirement that $f(0,X_0)=S_0$. Although the term structure given by Eq.(\ref{singularterm}) is singular at $t=0$, it serves as a valid flow function at $t=0$ as long as $f(0,X_0)=S_0$. The specific form of the flow function term structure is a modeling choice dependent on the relationship between the flow variable and the target variable. For example, if the flow variable is a Brownian motion $W_t$ and the target distribution is log-normal, a linear interpolation in the logarithm of the flow function is more appropriate. The martingale property of $S_t$ given in Eq.(\ref{fpde}) imposes the following relation between the flow function and the drift function:
\begin{align}
\label{timelineardrift}
\mu(t,x) = -\frac{\partial_t f(t,x)+ \frac{1}{2}\partial_x^2 f(t,x)}{\partial_x f(t,x)},
\end{align}
where the time derivative $\partial_t {f}(t,x)\triangleq \frac{\partial f}{\partial t}(t,x)$ can be evaluated according to the specified time-dependence of $f(t,\cdot)$. The following fixed-point equation produces such a solution when the iteration converges (Algorithm \ref{continuous_flow_fixed_point}):

\begin{algorithm}[H]
\caption{Continuous flow fixed-point iteration for $t\in[T_{i},T_{i+1})$}\label{continuous_flow_fixed_point}
\begin{algorithmic}
\State \textbf{Input}: $F_{X_{T_i}}(x)$ and $f(T_i,x)$.
\State Start with an initial guess of $\mu(x)$.
\Repeat
\State Solve the forward equation (\ref{ppde}) to obtain $F_{X_{T_{i+1}}}(x)$.
\State Set the flow function $f(T_{i+1},x)$ at maturity $T_{i+1}$ by Eq.(\ref{quantilematching}).
\State Interpolate the flow function, for example, according to Eq.(\ref{singularterm}).
\State Update drift function by Eq.(\ref{timelineardrift}).
\Until{$f(T_{i+1},x)$ converge.}
\end{algorithmic}
\end{algorithm}

Subject to the condition that the fixed-point iteration converges, the resulting solution produces a continuous local volatility function across the marginal maturities. In a Monte Carlo simulation, there is no need for matching the flow variable over successive forward periods by the continuity condition such as Eq.(\ref{continuity1}) or Eq.(\ref{continuity2}). However, the term structure of the resulting local volatility function is often oscillatory. Consider an underlying volatility model $\sigma_t$ and the construction of a local volatility model $\sigma_{\text{loc}}(t,\cdot)$ matching the marginal distributions at $T_i$ as well as the corresponding variance swaps $\int_{T_{i-1}}^{T_{i}}\mathbb{E}[\sigma_t^2] dt$, for $i=1,2,\cdots,n$. Unless $\mathbb{E}[\sigma_{\text{loc}}^2(t,\cdot)]$ matches up with $\mathbb{E}[\sigma_t^2]$ exactly for $\forall t\in[T_0,T_1]$, an undershooting (overshooting) at $t=T_1$ inevitably leads to an overshooting (undershooting) at $t=T_2$ and an oscillatory term structure over $T_2,T_3,\cdots,T_n$.

\section{Examples}
\label{example}

To illustrate the local volatility constructions described previously, we consider two sets of test cases: one case with a synthetic family of double exponential marginal distributions and another with market options data. In the latter case, a discrete set of option prices is observed for each maturity. Before applying the Markov-functional method to interpolate volatility in the maturity dimension, we perform a parametric fitting of the implied volatility skews, hence the risk-neutral distributions, for each market maturity. The method of skew fitting is described in Appendix \ref{skew_fitting}.

\subsection{Double Exponential Distributions}

We first consider the family of double negative exponential densities for the target variable $S_t=Y_t$:
\begin{align}
\label{doublexponentialdensity}
p(t,y) = \frac{\lambda(t)}{2}e^{-\lambda(t)|y|},
\end{align}
where $\lambda(t)$ is a positive decreasing function of time $t$. For example, we choose $\lambda(t)=\frac{1}{\sqrt{t}}$ as in Madan and Yor, 2002 \cite{madan2002}. In this model, the prices of vanilla options, the local volatility function (Bachelier-type), the marginal CDF and its inverse are all known in closed form.

\begin{figure}[!ht]
\centering
\includegraphics[scale=0.34]{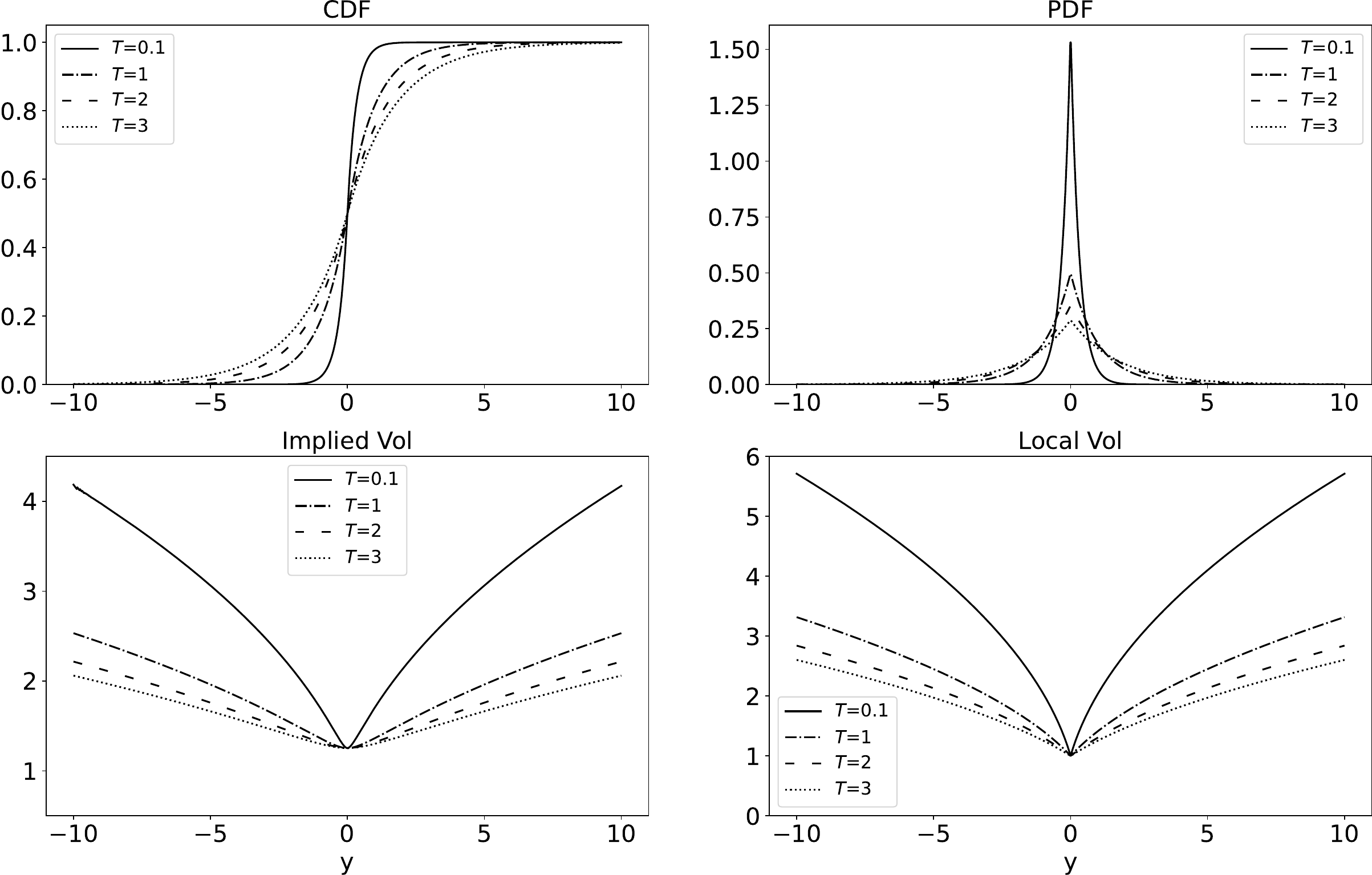}
\caption{The CDF, PDF, implied volatility and Bachelier local volatility of the double negative exponential distribution with $\lambda(t)=\frac{1}{\sqrt{t}}$ at $t=0.1,1,2$.}
\end{figure}

\begin{itemize}
\item The call option price expiring at $T$ and struck at $K$
\begin{align}
C(T,K) \triangleq \mathbb{E}\left[(Y_T-K)^+\right] = (-K)^+ + \frac{e^{-\lambda |K|}}{2\lambda},
\end{align}
is a decreasing function of $\lambda(T)$. So, the densities given by Eq.(\ref{doublexponentialdensity}) are in convex order with increasing $T$.
\item Denote $\dot{\lambda}\triangleq \frac{d\lambda}{dt}<0$, the Bachelier-type local volatility function
\begin{align}
\label{doubleexponentiallocalvol}
\sigma^2_{\text{loc}}(t,y) = 2\left(\frac{\partial C}{\partial T}\right) \left(\frac{\partial^2 C}{\partial K^2}\right)^{-1}\Big|_{K=y,T=t}
= \frac{2(1+\lambda|y|)(-\dot{\lambda})}{\lambda^3}.
\end{align} 
The corresponding flow function is given by Eq.(\ref{flowfunctionode}):
\begin{align}
\label{doubleexponentialflow}
f(t,x) = \frac{1}{\lambda}\left[ \left( 1+\sqrt{\frac{-\dot{\lambda}}{2\lambda}}x \right)^2-1 \right].
\end{align}
\item The CDF and its inverse are:
\begin{align}
& F(t,y) \triangleq \mathbb{P}(Y_t\le y) = \begin{cases}
1 - \frac{1}{2}e^{-\lambda y}, & y\ge 0,\\
\frac{1}{2}e^{\lambda y}, & y<0.
\end{cases},\\
& F^{-1}(t,q) = \begin{cases}
-\frac{\log 2(1-q)}{\lambda}, & q\ge\frac{1}{2},\\
\frac{\log 2q}{\lambda}, & q<\frac{1}{2}.
\end{cases}
\end{align}
\end{itemize}

To test the various volatility interpolation methods for this synthetic case, we choose to observe a discrete set of marginals at $T_1=0.1$, $T_2=1$, $T_3=2$, $T_4=3$.

\subsubsection*{The Bass and, the Conze and Henry-Labord\`{e}re Interpolations}

\begin{figure}[!ht]
\centering
\includegraphics[scale=0.235]{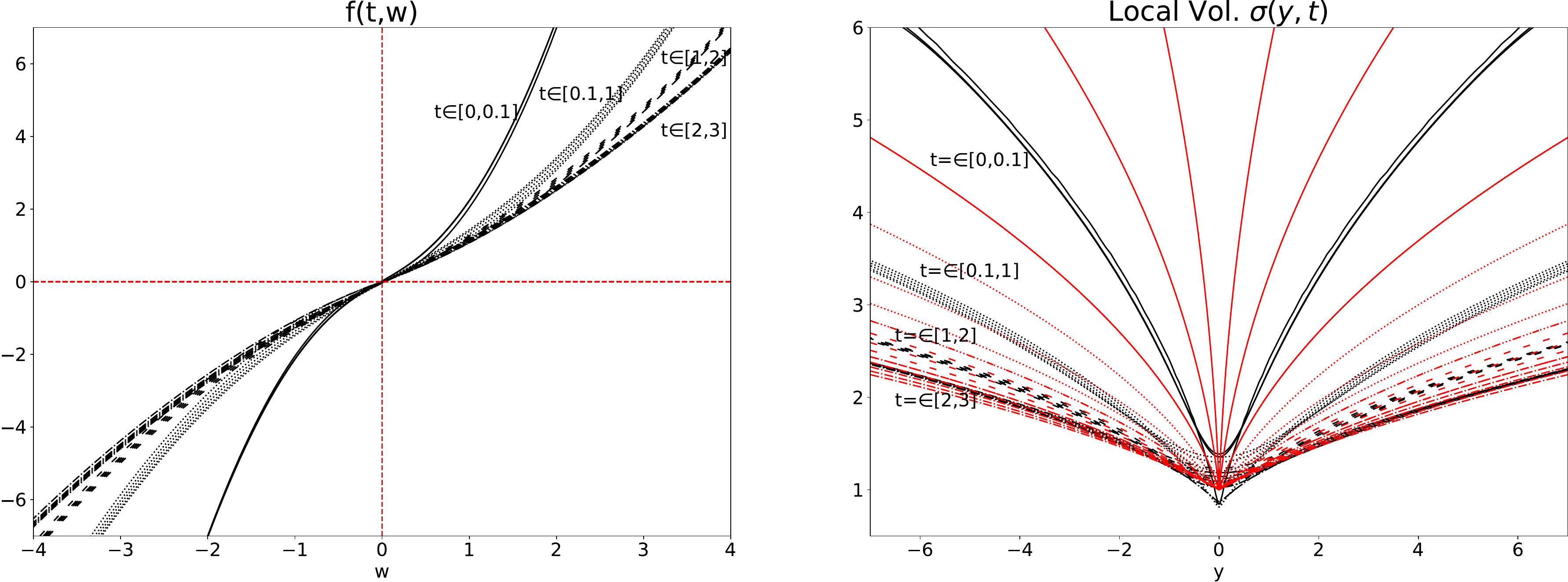}
\caption{Left panel: the flow function $f(t,w)$ of Bass for $t\in[0,0.1]$ and Conze and Henry-Labord\`{e}re for $t\in [0.1,3]$. Right panel: the local volatility function $\sigma_{\text{loc}}(t,y)$ of Bass for $t\in[0,0.1]$ and Conze and Henry-Labord\`{e}re for $t\in [0.1,3]$, where the original model Eq.(\ref{doubleexponentiallocalvol}) are shown in red.}
\label{bass_phl_flow}
\end{figure}

\begin{figure}[!ht]
\centering
\includegraphics[scale=0.235]{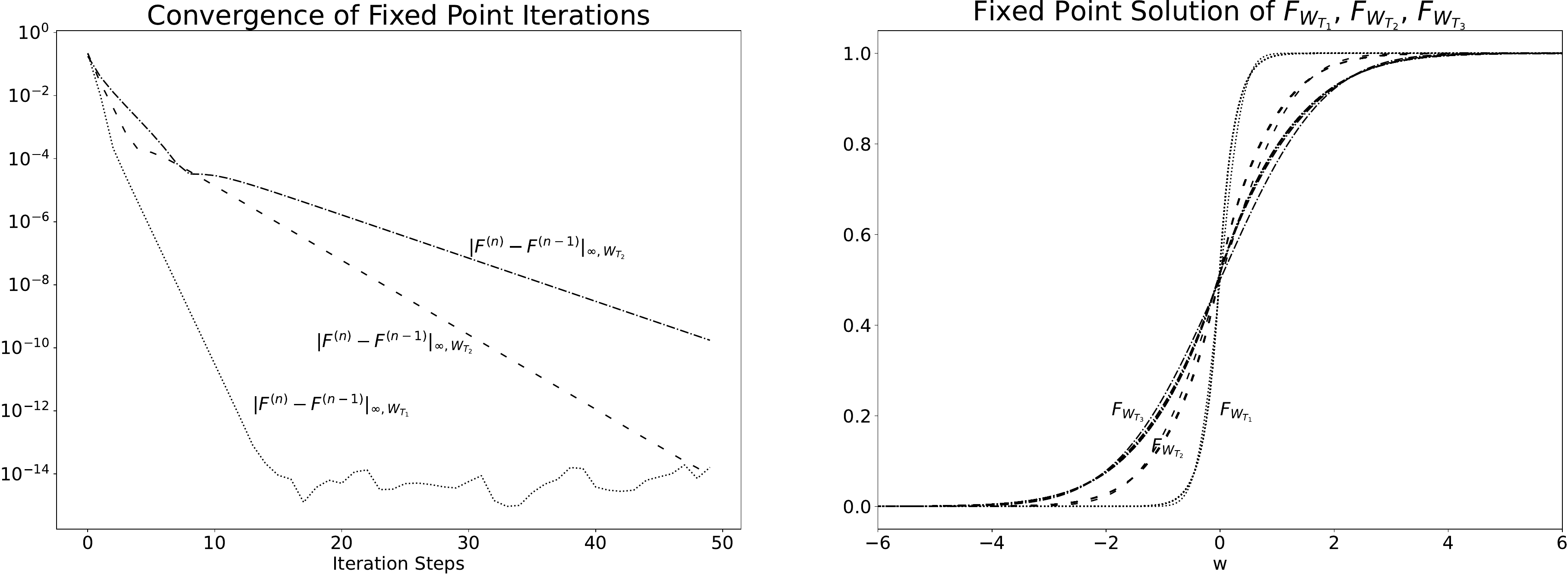}
\caption{Left panel: the convergence of the norm of the successive differences $|F^{(n)}-F^{(n-1)}|$ of the flow variable CDF $F_{W_{T_1}}$, $F_{W_{T_2}}$ and $F_{W_{T_3}}$.  Right panel: The flow variable CDF $F_{W_{T_1}}$, $F_{W_{T_2}}$ and $F_{W_{T_3}}$, where different lines of the same style corresponds to different iterations.}
\label{bass_phl_convergence}
\end{figure}

For the first forward period $t\in[0,T_1]$, the method of Bass 1983 \cite{bass1983} is applied, while for the subsequent periods $t\in[T_1, T_2]$, $[T_2,T_3]$ and $[T_3,T_4]$, the method of Conze and Henry-Labord\`{e}re 2022 \cite{phl2022} is applied. The resulting flow functions and the local volatility functions evaluated at a series of time points $t\in\{0.001, 0.01, 0.1, \cdots, 3.0\}$ are shown in Figure (\ref{bass_phl_flow}). We used fast Fourier transform (FFT) to evaluate the convolution integrals in Eq.(\ref{bass_quadrature_1}), (\ref{phl_quadrature_1}) and (\ref{phl_quadrature_2}). The resulting CDFs of the flow variable at $T_i$, $i=1,2,3,4$ are interpolated on a grid. The fixed-point equations converge rapidly, as shown in Figure (\ref{bass_phl_convergence}).

\subsubsection*{Time-Homogeneous Interpolation}
\begin{figure}[!ht]
\centering
\includegraphics[scale=0.235]{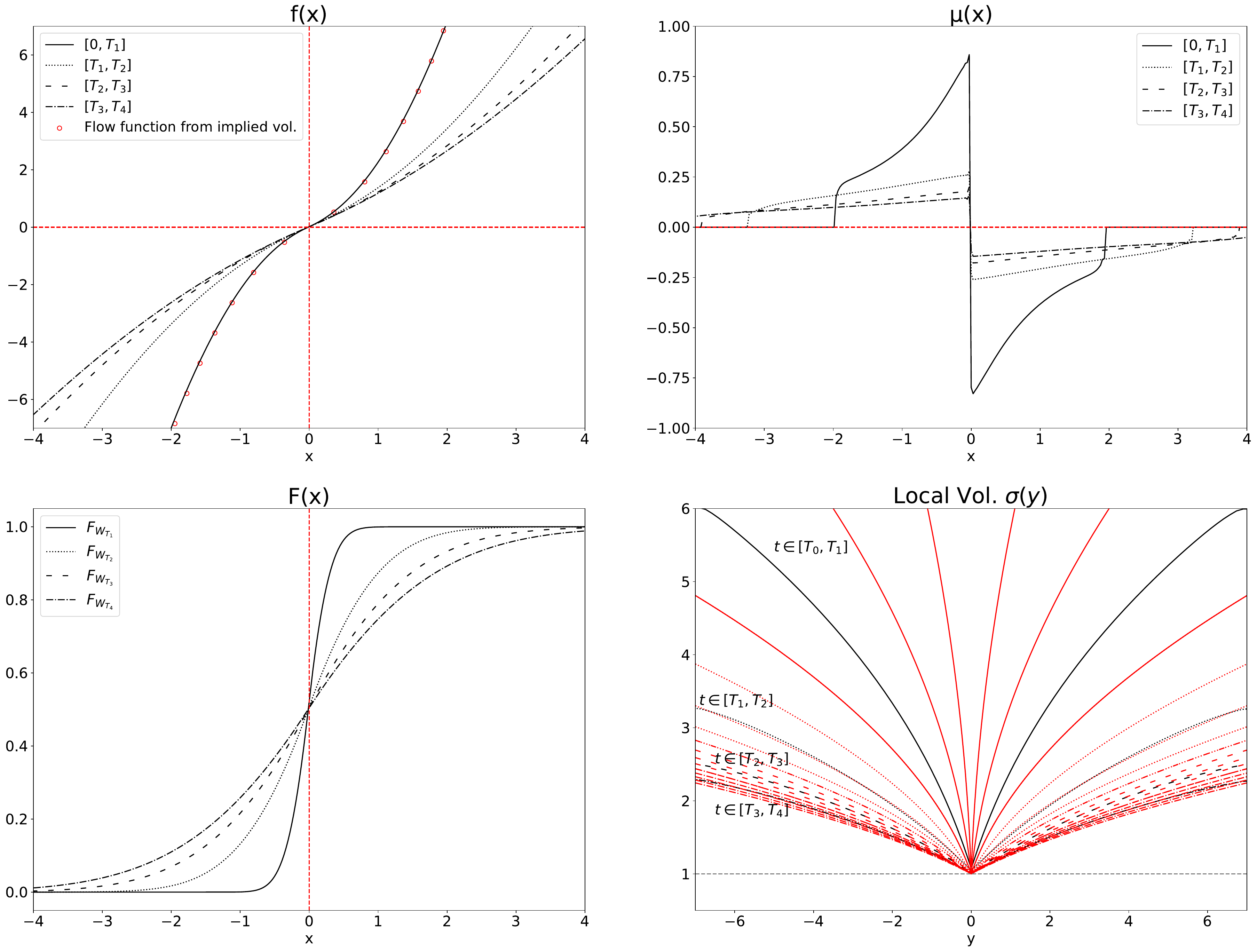}
\caption{Time-homogeneous local volatility model per forward period $t\in[0,T_1]$, $[T_1, T_2]$, $[T_2,T_3]$ and $[T_3,T_4]$: flow functions (top-left), drift functions (top-right), flow variable CDF (bottom-left) and local volatility functions (bottom-right), where the original model Eq.(\ref{doubleexponentiallocalvol}) are shown in red lines. The red circles in the top-left panel are short-term flow functions calculated from its relationship with the implied volatility skew Eq.(\ref{bbfrelation}).}
\label{krylov_flow}
\end{figure}

We apply Algorithm \ref{krylov_fixed_point_1} to the first forward period $t\in[0,T_1]$ and Algorithm \ref{krylov_fixed_point_2} to the subsequent periods $t\in[T_1, T_2]$, $[T_2,T_3]$ and $[T_3,T_4]$. The resulting flow functions and the local volatility functions have a step-wise term structure in the time dimension: $f(T_i,x)$ and $\sigma_{\text{loc}}(T_i,x)$, for $i=1,2,3$, as shown in Figure (\ref{krylov_flow}).

\begin{figure}[!ht]
\centering
\includegraphics[scale=0.235]{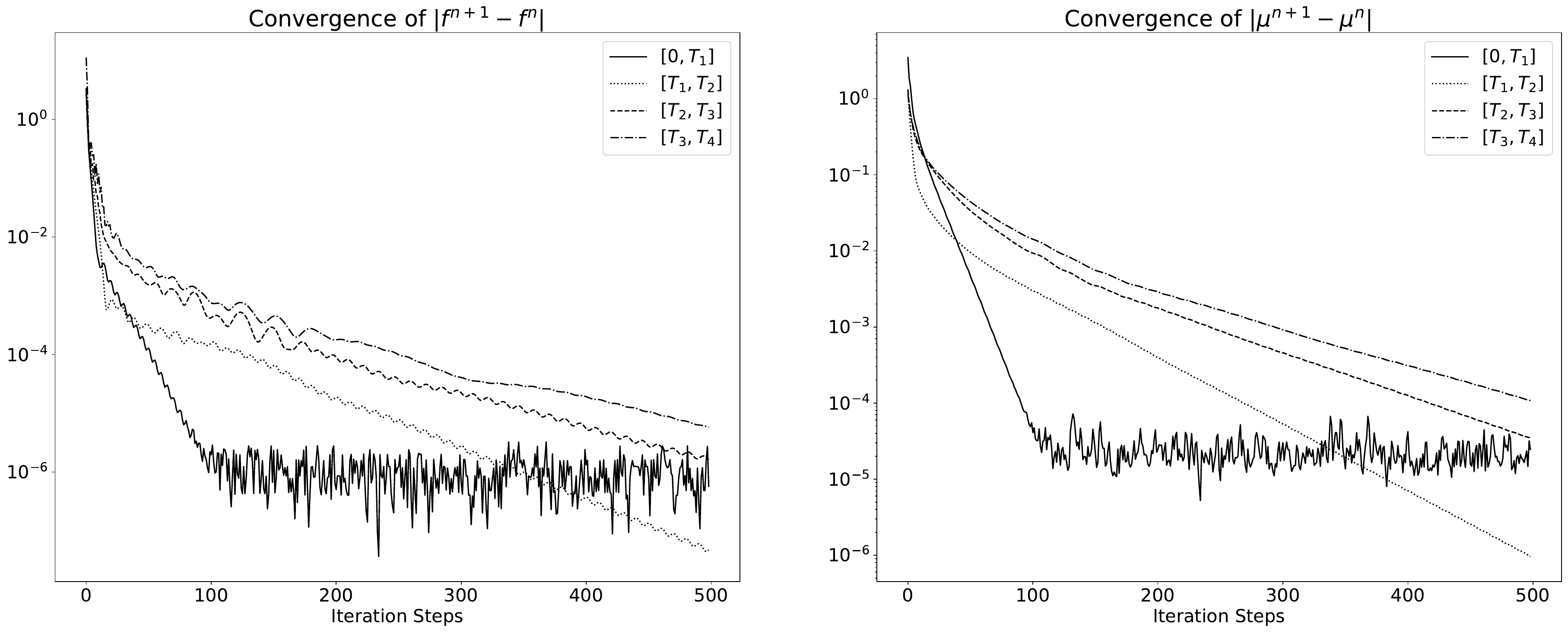}
\caption{Time-homogeneous local volatility model per forward period $t\in[0,T_1]$, $[T_1, T_2]$, $[T_2,T_3]$ and $[T_3,T_4]$: the convergence of the norm of the success differences $|f^{(n)}-f^{(n-1)}|$ of the flow function (left panel) and $|\mu^{(n)}-\mu^{(n-1)}|$ of the drift function (right panel).}
\label{krylov_convergence}
\end{figure}

\begin{itemize}
\item To perform the quantile matching $F^{-1}_{\nu_i}\circ F_{X_{T_i}}(x)$ in the regime where the probability density is vanishing, we extrapolate the flow function linearly beyond a certain threshold: $|f(x)|>y_{\text{max}}$, effectively setting the drift function $\mu(x)=0$ due to Eq.(\ref{krylovdrift}), as shown in the top-right panel of Figure (\ref{krylov_flow}).
\item To remove the ambiguity in the flow variable and the flow function, we enforce that $\mathbb{E}[X_{T}]=0$. Numerically, the probability density $p_i$ of the flow variable is defined on a grid $x_i$, $i=1,2,\cdots,N$. We apply a multiplier $C_\pm$, at each step of the fixed-point iteration, to all $p_i$'s for which $\pm x_i>0$, and $\frac{1}{2}(C_+ + C_-)$ to the probability $p_{i_0}$ for $x_{i_0}=0$. This leads to the following linear equations for $C_\pm$:
\begin{align}
\begin{cases}
C_+ \left(\sum_{x_i>0}p_i + \frac{1}{2} p_{i_0}\right) + C_-\left(\sum_{x_i<0}p_i + \frac{1}{2} p_{i_0}\right) = 1, \\
C_+ \sum_{x_i>0}p_i x_i + C_-\sum_{x_i<0}p_i x_i = 0. 
\end{cases}
\end{align}
\item We used a $(100,500)$ grid of $(t,x)$ to solve the forward equation and represent the flow functions, drift functions, etc. The range of the flow variable $x$ is determined according to a range of the target variable $y\in[-7,7]$. The successive differences $|f^{(n)}-f^{(n-1)}|$ and $|\mu^{(n)}-\mu^{(n-1)}|$ are measured at each iteration step of the fixed-point equation. The convergence, shown in Figure (\ref{krylov_convergence}), is still geometric at a lower rate than the method of Bass 1983 \cite{bass1983}, Conze and Henry-Labord\`{e}re 2022 \cite{phl2022}, shown in Figure (\ref{bass_phl_convergence}). The rates of convergence $r$ for the flow functions $|f^{(n)}-f^{(n-1)}|\sim r^n$ are reported in Table (\ref{rate_of_convergence_table}).
\begin{table}[!ht]
    \centering
    \begin{tabular}{c|r|r}
    \hline
    Forward Period & Range of Iteration Steps & Rate of Convergence $r$  \\
    \hline
    $[T_0,T_1]$    & $50 \le n \le 100 $ & 0.912 \\
    $[T_1,T_2]$    & $100 \le n \le 500 $ & 0.980 \\
    $[T_2,T_3]$    & $100 \le n \le 500 $ & 0.986 \\
    $[T_3,T_4]$    & $100 \le n \le 500 $ & 0.986 \\
    \hline
    \end{tabular}
    \caption{The rates of convergence $r$ for the step-wise time-homogeneous flow functions $|f^{(n)}-f^{(n-1)}|\sim r^n$ per forward period $t\in[0,T_1]$, $[T_1, T_2]$, $[T_2,T_3]$ and $[T_3,T_4]$. The second column represents the range of iteration steps that are used to estimate the rate of convergence.}
    \label{rate_of_convergence_table}
\end{table}


\end{itemize}

\subsubsection*{Continuous Interpolation in Time Dimension}

\begin{figure}[!ht]
\centering
\includegraphics[scale=0.235]{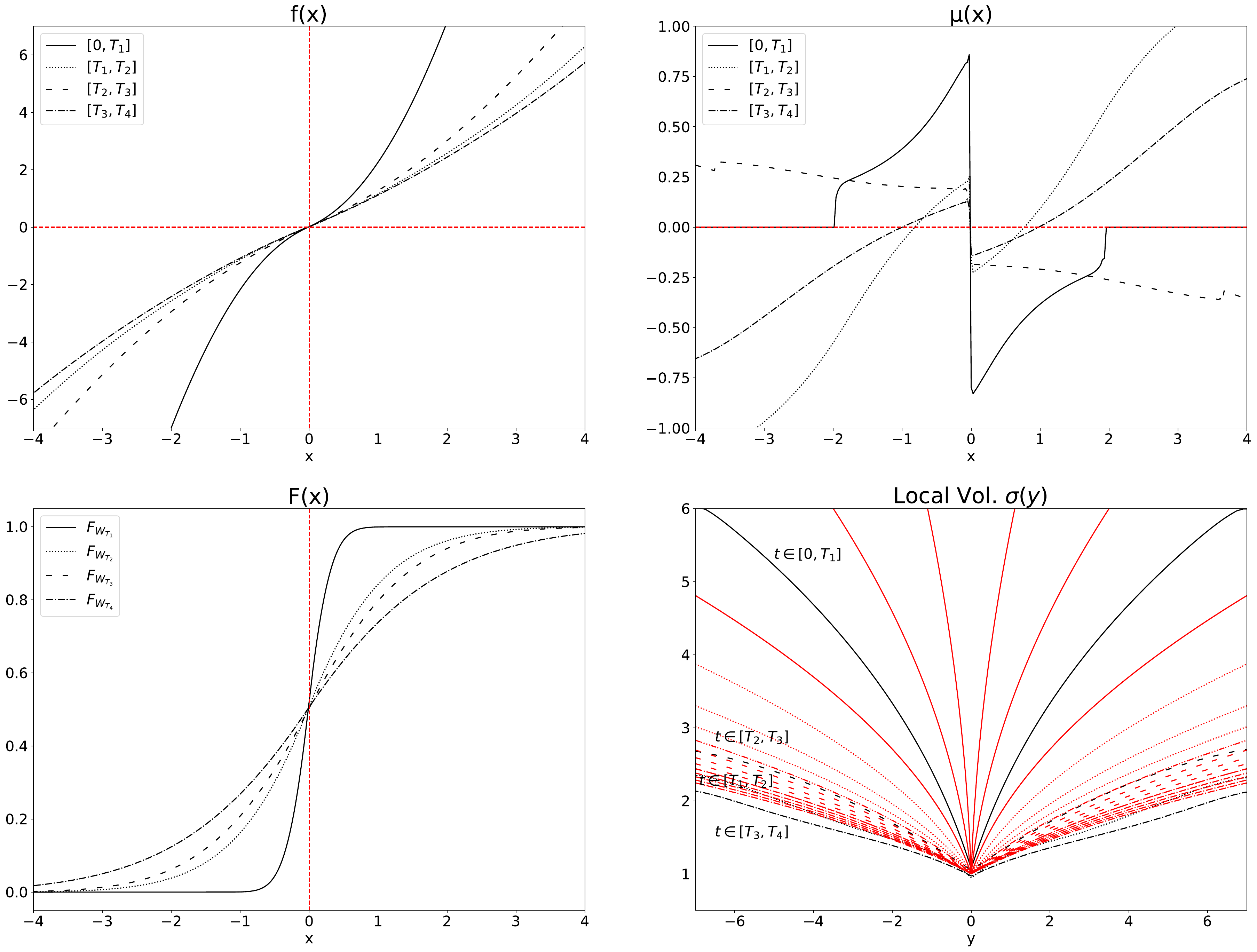}
\caption{Continuous local volatility model $t\in[0,T_1]$, $[T_1,T_2]$, $[T_2,T_3]$ and $[T_3,T_4]$ where the first forward period $[0,T_1]$ is interpolated by a Krylov-type time-homogeneous model, same as that of Figure (\ref{krylov_flow}): flow functions (top-left), drift functions (top-right), flow variable CDF (bottom-left) and local volatility functions (bottom-right), where the original model Eq.(\ref{doubleexponentiallocalvol}) are shown in red.}
\label{continuous_flow}
\end{figure}

\begin{figure}[!ht]
\centering
\includegraphics[scale=0.235]{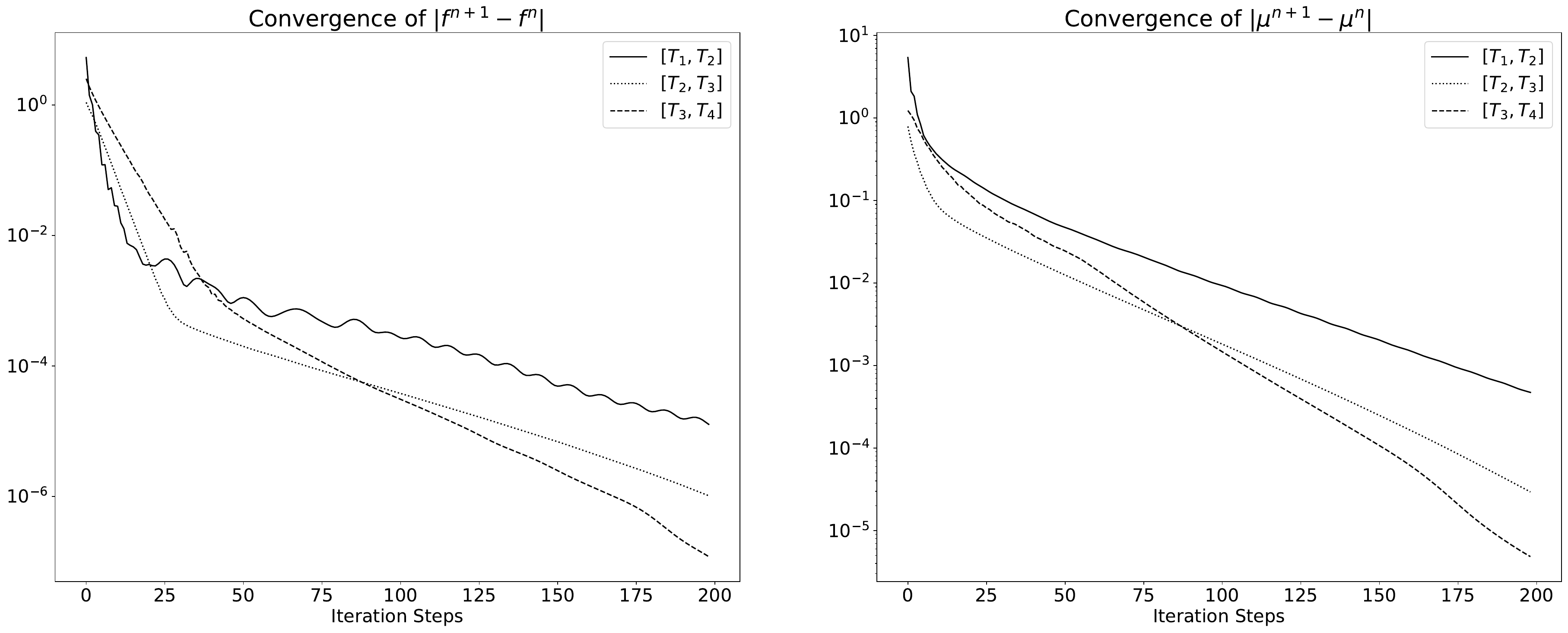}
\caption{Continuous local volatility model per forward period  $t\in[0,T_1]$, $[T_1,T_2]$, $[T_2,T_3]$ and $[T_3,T_4]$ where the first forward period $[0,T_1]$ is interpolated by a Krylov-type time-homogeneous model: the convergence of the norm of the success differences $|f^{(n)}-f^{(n-1)}|$ of the flow function (left panel) and $|\mu^{(n)}-\mu^{(n-1)}|$ of the drift function (right panel).}
\label{continuous_convergence}
\end{figure}

We proceed to construct a local volatility surface that is continuous in the time direction, specifically, across the marginal maturities. As mentioned in Section (\ref{continuous_section}), the flow function at the origin of time $f(0,\cdot)$ is undetermined. In fact, in the example of the double negative exponential distribution under consideration, the local volatility function given by Eq.(\ref{doubleexponentiallocalvol}) with $\lambda(T)=\frac{1}{\sqrt{T}}$:
\begin{align}
\label{doubleexponentiallocalvolsurface}
\sigma^2_{\text{loc}}(t,y) = 1 + \frac{|y|}{\sqrt{t}}
\end{align}
the corresponding flow function given by Eq.(\ref{doubleexponentialflow})
\begin{align}
f(t,x) = \text{sign}(x)\cdot \frac{x^2}{4\sqrt{t}} + x,
\end{align}
are singular at $t=0$. Instead of postulating a form of the flow function at the origin of time $f(0,\cdot)$, we apply Algorithm \ref{krylov_fixed_point_1} to construct a time-homogeneous local volatility model for the first forward period $[0,T_1]$ and Algorithm \ref{continuous_flow_fixed_point} to perform the continuous flow interpolation for the remaining periods. We choose the term structure specified by Eq.(\ref{singularterm}). The resulting flow functions and local volatility functions are shown in Figure (\ref{continuous_flow}). We used the same $(100,500)$ grid of $(t,x)$ as in the time-homogeneous case to solve the forward equation and represent the flow functions, drift functions, and so forth. The successive differences $|F^{(n)}-F^{(n-1)}|$ and $|\mu^{(n)}-\mu^{(n-1)}|$ are measured at each iteration step of the fixed-point equation. The convergence is geometric, as shown in Figure (\ref{continuous_convergence}).

\begin{figure}[!ht]
\centering
\includegraphics[scale=0.235]{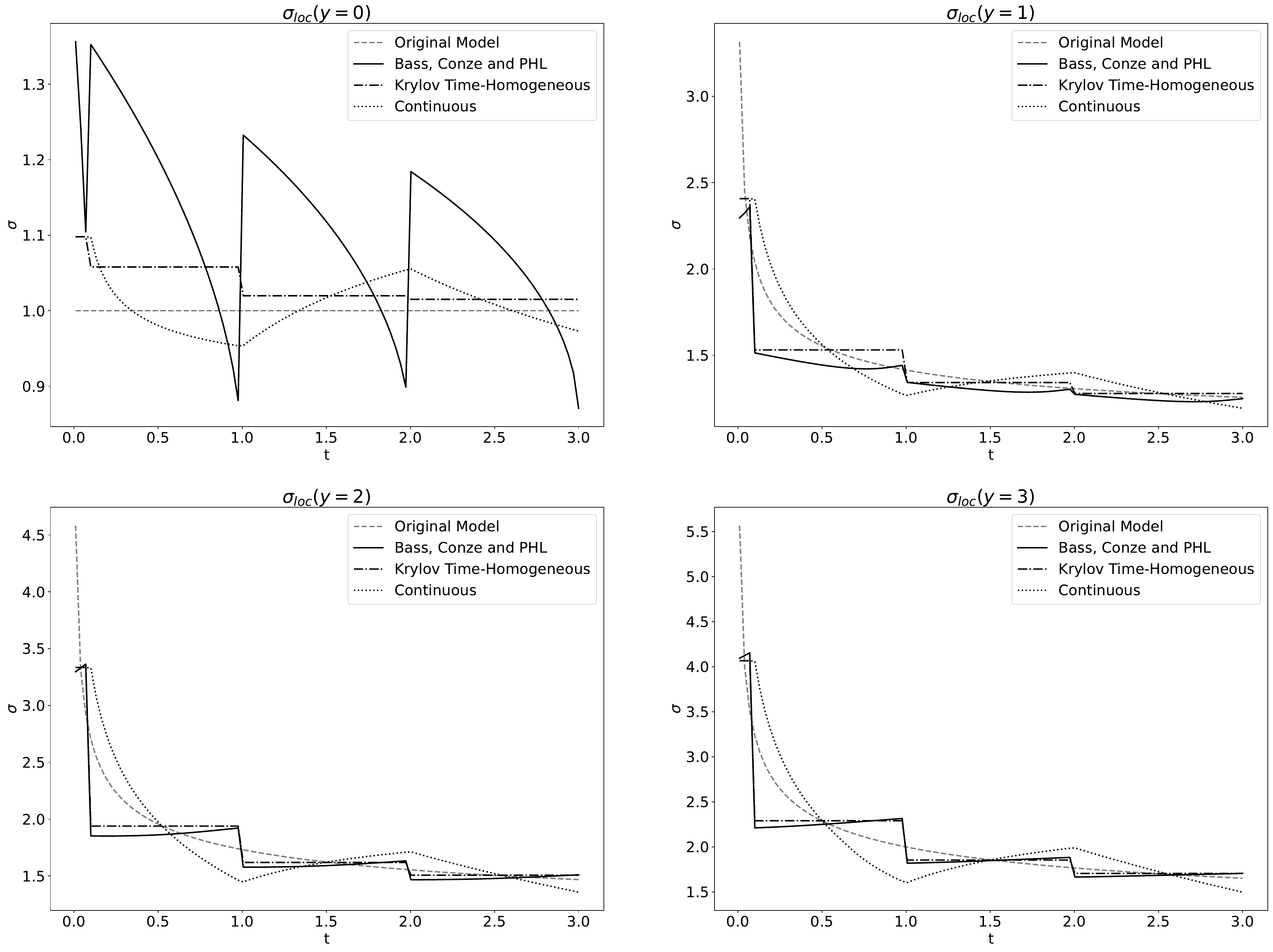}
\caption{The term structures in the time dimension of the local volatility surface of the continuum model and those generated by interpolations, at $y=0,1,2,3$.}
\label{double_exponential_compare}
\end{figure}

\begin{figure}[!ht]
\centering
\includegraphics[scale=0.45]{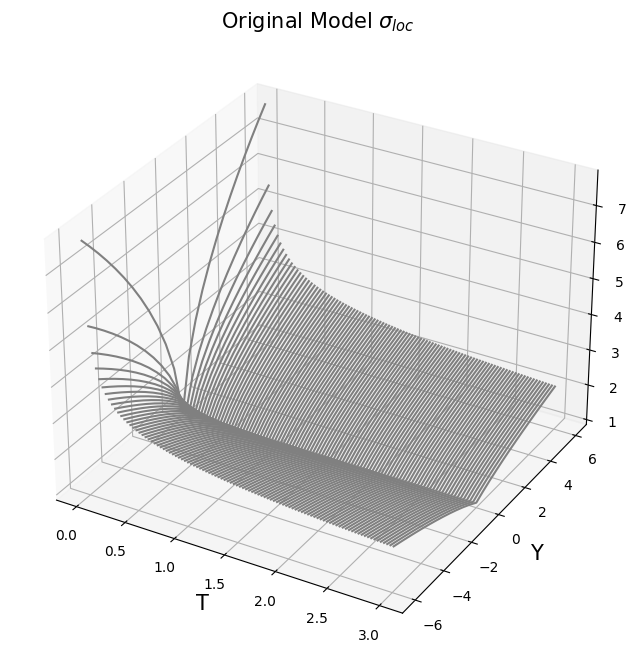}
\includegraphics[scale=0.45]{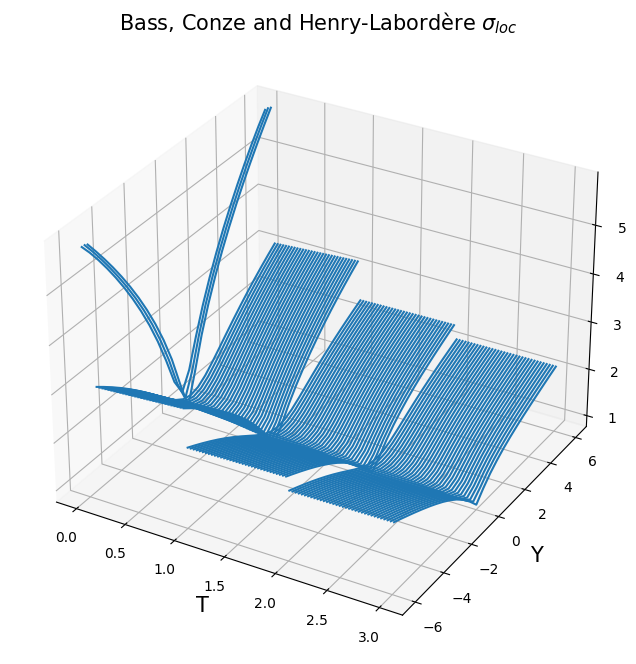}\\
\includegraphics[scale=0.45]{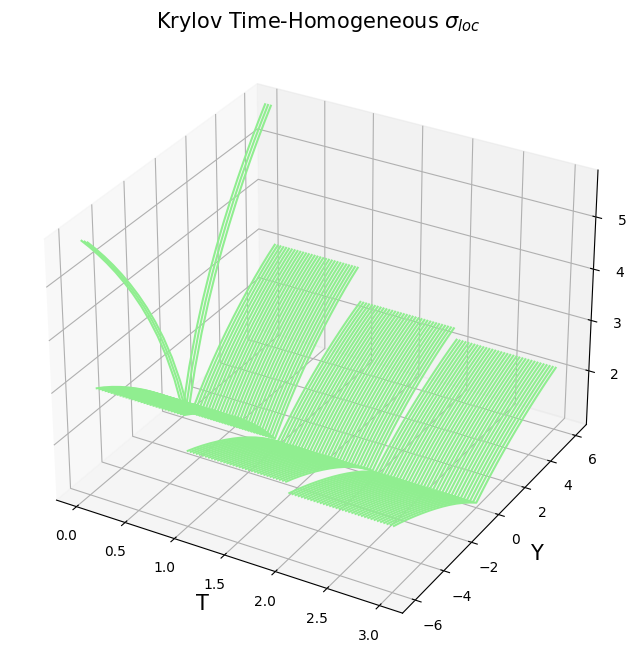}
\includegraphics[scale=0.45]{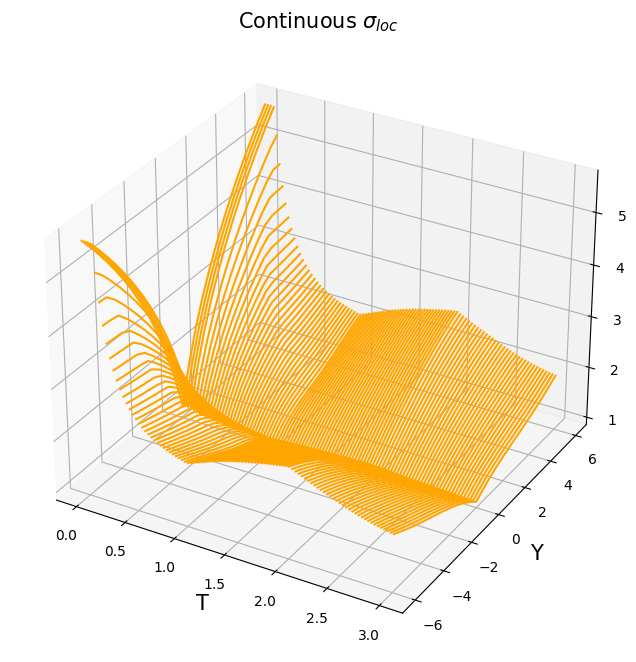}
\caption{The local volatility surfaces derived from the continuum model (top-left), interpolated from a discrete set of marginals at $t\in\{0.1, 1, 2, 3\}$ by Bass, Conze and Henry-Labord\`{e}re (top-right), Krylov time-homogeneous (bottom-left), and a continuous scheme (bottom-right). }
\label{double_exponential_surface_compare}
\end{figure}

In Figure (\ref{double_exponential_compare}), we compare the local volatility term structures of the continuum model (dashed lines) and those generated by interpolation schemes. The interpolated ATM ($y=0$) term structure given by Bass, and Conze and Henry-Labord\`{e}re (solid lines) exhibits the expected discontinuities across the marginal maturities and an unrealistic oscillation. The oscillation is much reduced for $y=1,2,3$. The time-homogeneous model (dash-dotted lines) generates the most parsimonious, monotonic but discontinuous, term structure. Because the continuum model given by Eq.(\ref{doubleexponentiallocalvolsurface}) is singular at $t=0$ for $y\neq 0$, the continuous interpolation (dotted lines), or any other model with a finite local volatility at $t=0$, overshoots at the beginning of the second period $[T_1,T_2]$, forcing it to undershoot at the end of the same period, to match the overall volatility level (or equivalently total variance) of the continuum model. This explains the alternating slopes of the term structure given by the continuous model, characteristic of a calibrated local volatility surface. Finally, the local volatility surface of the continuum model Eq.(\ref{doubleexponentiallocalvolsurface}) and those generated by the interpolating a discrete set of marginals at $t\in\{0.1, 1, 2, 3\}$ are shown in Figure (\ref{double_exponential_surface_compare}).

\subsection{Market Options Data: JPM, as of February 15, 2023}
\label{jpmsection}

\begin{figure}[!ht]

\centering
\includegraphics[scale=0.35]{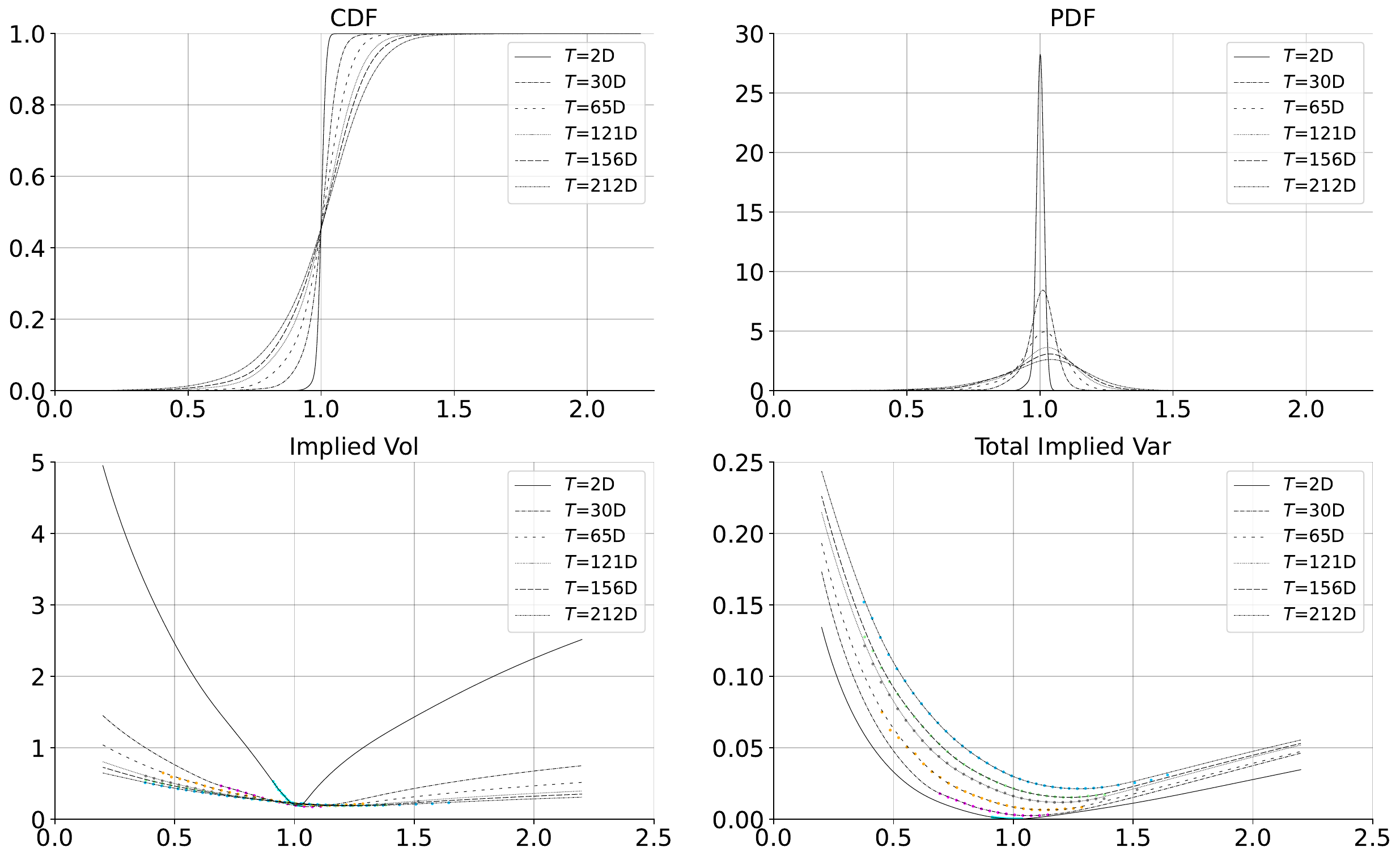}
\caption{JPMorgan options data (points) and the fit (lines) by mixed lognormal densities for each maturity successively, according to Algorithm \ref{mln_ah}.}
\label{jpm_data}
\end{figure}

The second test case we consider is the market options data of JPMorgan on February 15, 2023, shown in Figure (\ref{jpm_data}). For each strike, a de-Americanization procedure is performed to compute a Black implied volatility that matches up with the corresponding American option price. Also shown in Figure (\ref{jpm_data}) is the parametric fitting of the implied volatility skew at each maturity to a 4-mode mixed lognormal (MLN) risk neutral density. The procedure to avoid calendar spread arbitrage in the slice-by-slice fitting of the implied skews is detailed in Appendix \ref{skew_fitting}. With the calibrated target distributions at $T\in\{2, 30, 65, 121, 156, 212\}$ in units of days, we report the results of the step-wise time-homogeneous volatility interpolation algorithms in Section \ref{krylov_section}. For brevity, the results for the continuous Markov function described in Section \ref{continuous_section} are not reported. Constructing a continuous Markov function by interpolating the snapshots at $T_i$, $i=1,2,\cdots,n$, generally produces an oscillatory term structure --- an issue made worse by the irregularities in the market options data. To fix the scale, the target variable $S_t$, the JPMorgan price, as well as the option strikes are normalized by the forward. The initial value of the target value $S_0=1$. We choose the flow variable to have an initial condition $X_0=0$. 


\subsubsection*{Time-Homogeneous Interpolation}

\begin{figure}[!ht]
\centering
\includegraphics[scale=0.24]{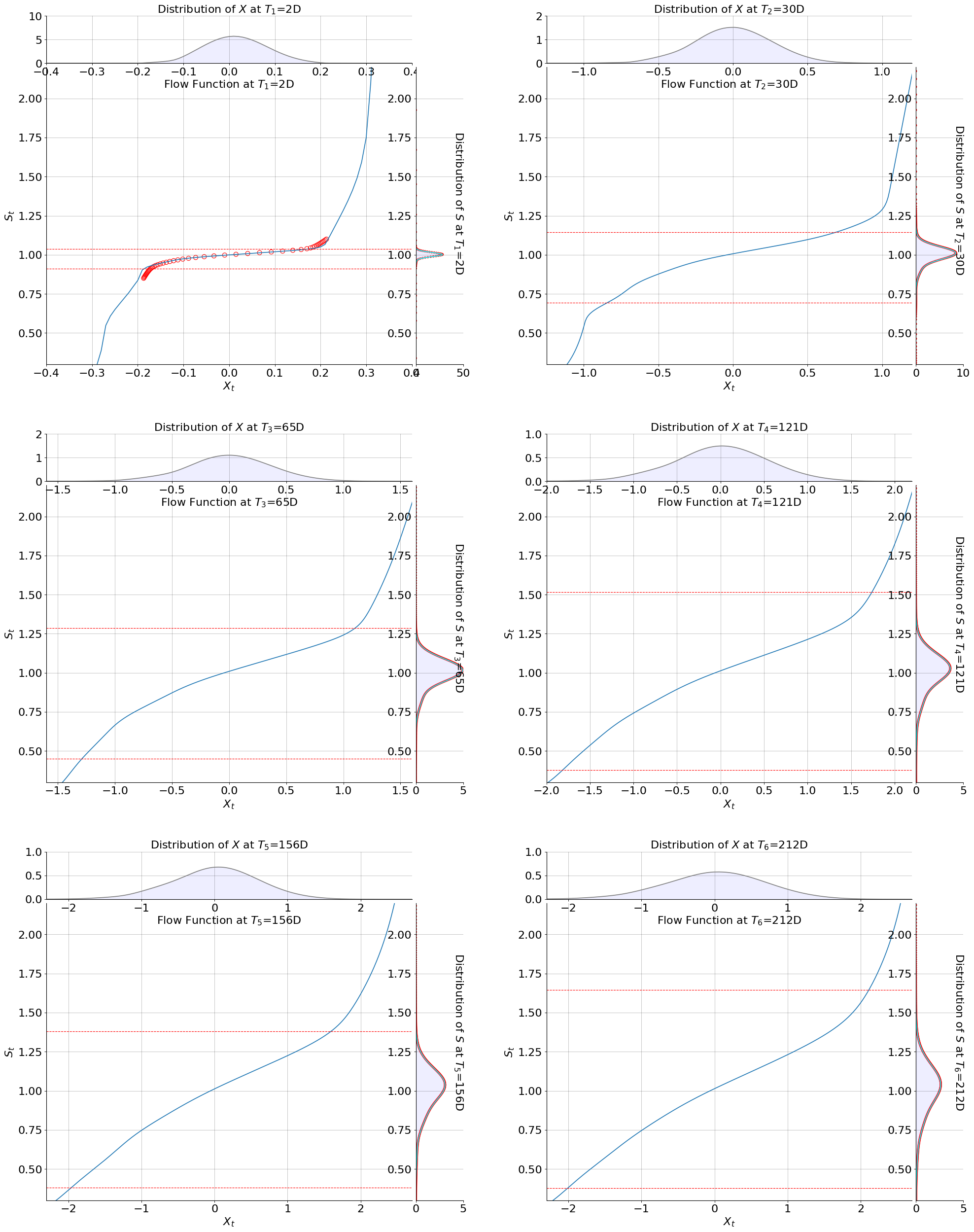}
\caption{Time-homogeneous flow functions for interpolating the JPMorgan options data at $T\in\{2, 30, 65, 121, 156, 212\}$ in units of days. The density functions of the flow variable and the target variable are plotted to the horizontal axis and the vertical axis, respectively. The red dashed lines on each chart mark the range of the market strikes for each maturity. The red circles on the top-left chart are short-term flow functions calculated from its relationship with the implied volatility skew Eq.(\ref{bbfrelation}).}
\label{jpm_krylov_flows}
\end{figure}

\begin{figure}[!ht]
\centering
\includegraphics[scale=0.24]{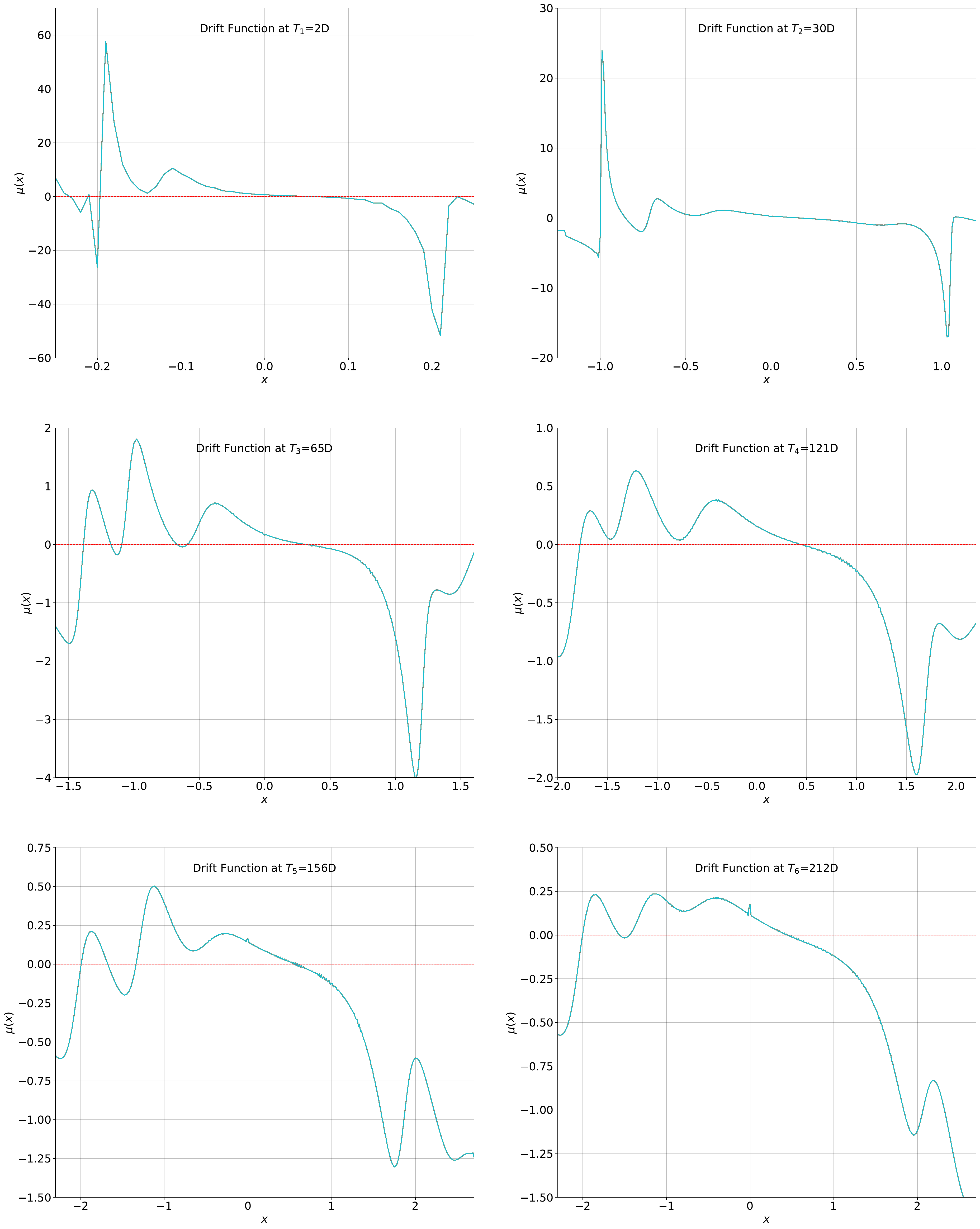}
\caption{Time-homogeneous drift functions $\mu(x)$ for interpolating the JPMorgan options data at $T\in\{2, 30, 65, 121, 156, 212\}$ in units of days. The resulting drift functions decrease in magnitudes over the periods $[T_{i-1}, T_i]$, but exhibit some oscillations due to the irregularities of the market option data.  To indicate the fixed-point convergence, the drift functions $\mu^{n}$ of the last 100 iterations of a total number of 500 iterations are drawn on top of each other in each plot, where no essential difference is seen.}
\label{jpm_krylov_drift}
\end{figure}

The step-wise time-homogeneous flow functions and drift functions are shown in Figure (\ref{jpm_krylov_flows}) and (\ref{jpm_krylov_drift}). The probability density profiles of the flow variable and the target variable are attached to the horizontal and the vertical axis, respectively. The probability density of the target variable given by the parametric MLN fitting is plotted (red line) on the vertical axis of each chart in Figure (\ref{jpm_krylov_flows}) to compare with the numerical solution (light blue area under the curve). Over the successive periods $[T_{i-1},T_{i})$, $i=1,2,3,4,5,6$, the time-homogeneous flow function $f_{[T_{i-1},T_{i})}(x)$ gets increasingly flattened. Correspondingly, the profile of the drift functions also decrease in magnitudes over the periods $[T_{i-1},T_{i})$. On each chart in Figure (\ref{jpm_krylov_flows}), the two horizontal dashed lines (in red) represent the boundary of the option strikes in the data. The flow functions and the corresponding local volatility functions beyond the strike boundary contains extrapolation artifacts in the MLN fitting of implied skews. For the shortest maturity in 2D, we also plot the flow function calculated using Eq.(\ref{bbfrelation}), in red circles. The results agree well with the flow function calculated by Algorithm \ref{krylov_fixed_point_1}.

\begin{figure}[!ht]
\centering
\includegraphics[scale=0.35]{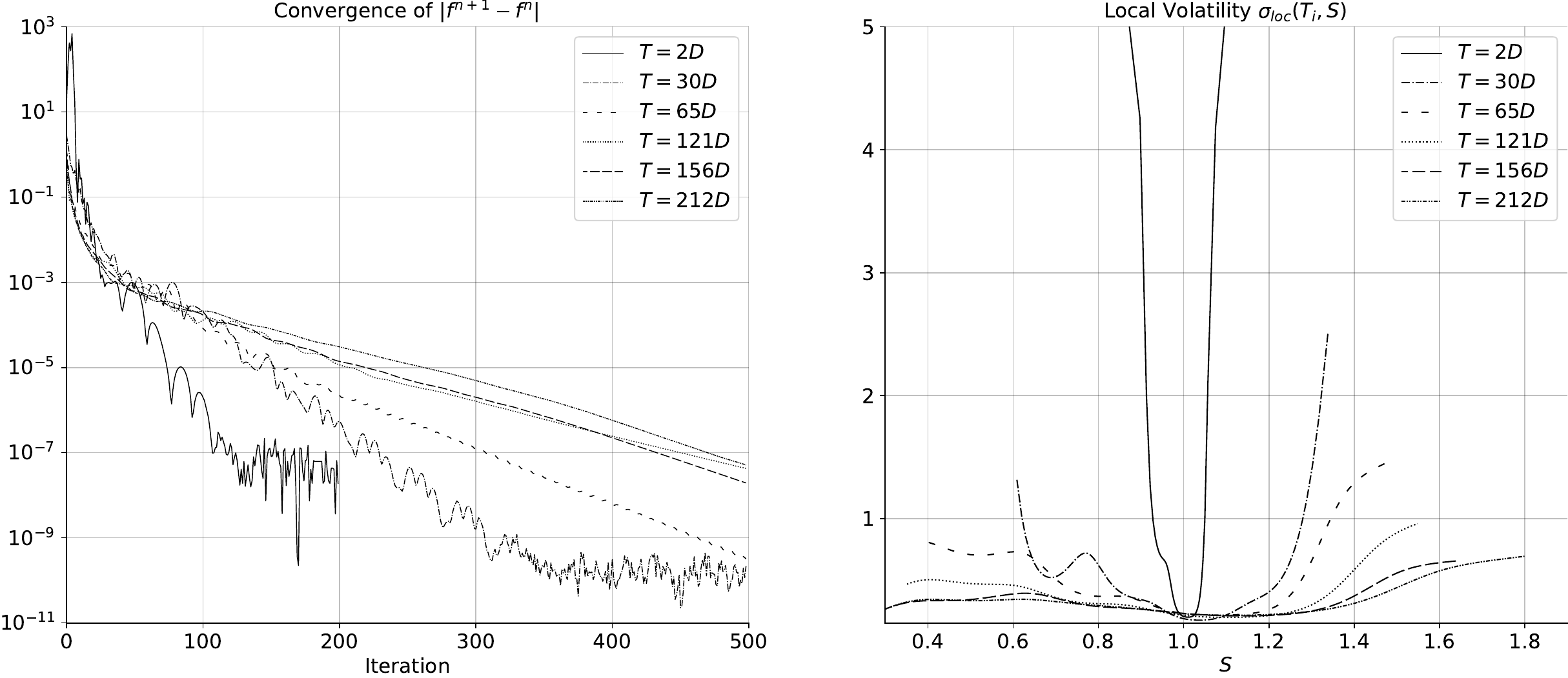}
\caption{Left panel: the convergence of the successive differences in the flow functions over fixed-point iterations. Right panel: the time-homogeneous local volatility functions interpolating the JPMorgan options data at $T\in\{2, 30, 65, 121, 156, 212\}$ in units of days, plotted in the range $[0.85\times K_{\min}, 1.15\times K_{\max}]$, where $K_{\min}$ and $K_{\max}$ denote the boundary of the observed strikes at each maturity.}
\label{jpm_krylov_localvol}
\end{figure}

The convergence of iterating the fixed-point equations is rapid, as shown in the left panel of Figure (\ref{jpm_krylov_localvol}). In the right panel, the time-homogeneous local volatility functions are calculated from a finite-difference approximation of Eq.(\ref{localvolandflow}) and plotted in the range $[0.85\times K_{\min}, 1.15\times K_{\max}]$, where $K_{\min}$ and $K_{\max}$ denote the boundary of the observed strikes at each maturity. In the current case of real market data, the underlying continuum model is unknown. Overall, the constructed local volatility function starts highly peaked near $T=0$ and rapidly decreases in magnitude as the time variable $T$ increases, similar to the synthetic case of the double exponential distributions shown in Figure (\ref{double_exponential_surface_compare}).
 
\newpage

\begin{figure}[!ht]
    \centering
    \includegraphics[scale=0.24]{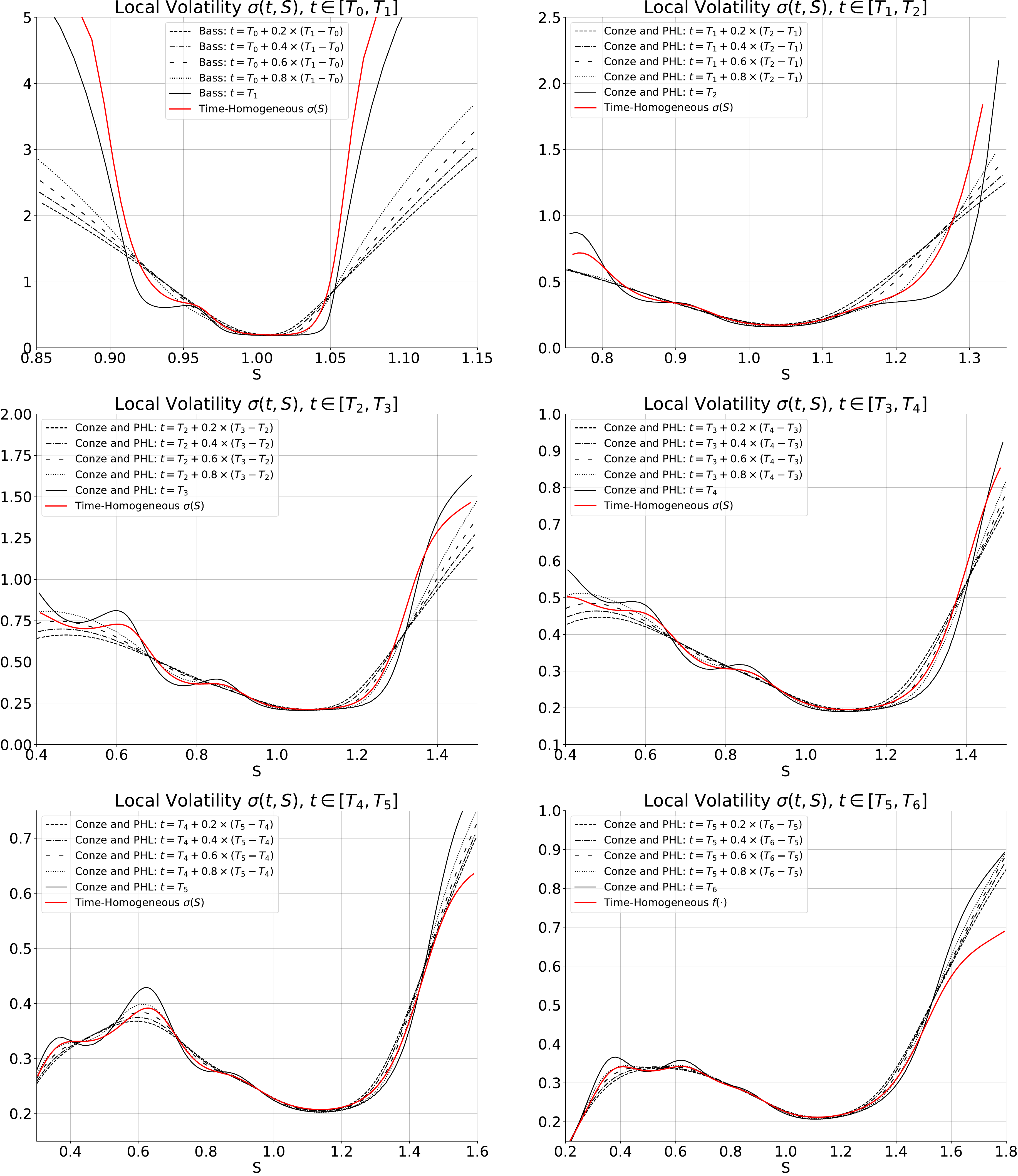}
    \caption{Time-homogeneous local volatility functions (red solid lines) and Bass (1983), Conze and Henry-Labord\`{e}re (2022) constructions (black lines) for interpolating the JPMorgan options data at $T\in\{2, 30, 65, 121, 156, 212\}$ in units of days. The same set of step-wise time-homogeneous local volatility functions are shown in Figure (\ref{jpm_krylov_localvol}) (right panel) in one graph.}
    \label{fig:jpm_compare_bass_krylov}
\end{figure}

In Figure (\ref{fig:jpm_compare_bass_krylov}), we plot the step-wise time-homogeneous local volatility constructions $\sigma(\cdot)$ and the constructions of Bass (1983) \cite{bass1983}, Conze and Henry-Labord\`{e}re (2022) \cite{phl2022} $\sigma(t,\cdot)$. Subject to some boundary effects at extreme strikes, the time-homogeneous constructions represent an average local volatility function over each successive period $[T_i,T_{i+1}]$, $i=0,1,2,3,4,5$.

\section{Summary}
\label{conclusion}

This paper develops an alternative perspective of viewing local volatility models --- the Markov functional model (MFM) \cite{bergomibook}. In this framework, a local volatility model can be specified in two equivalent ways: (1) the local volatility function $\sigma_{\text{loc}}(t,y)$; (2) the flow function $f(t,x)$ and the drift function $\mu(t,x)$. The linkage among the three functions, as well as the risk-neutral density $p(t,y)$ of the target variable, is shown in Figure (\ref{threewaylink}):

\begin{figure}[!ht]

\centering

\begin{tikzpicture}[font=\large]

\node[text width=1.5cm] (p0) at (1,0) {$p(t,y)$};
\node[text width=2cm] (p1) at (5,0) {$\sigma_{\text{loc}}(t,y)$};
\node[text width=1.5cm] (p2) at (9,0) {$f(t,x)$};
\node[text width=1.5cm] (p3) at (13,0) {$\mu(t,x)$};

\node[text width=2cm] at (3.5,0.3) {\footnotesize{\cite{dupire1994,derman1994}}};
\node[text width=2cm] at (7.7,0.3) {\footnotesize{Eq.(\ref{flowfunctionode})}};
\node[text width=2cm] at (11.5,0.3) {\footnotesize{Eq.(\ref{timelineardrift})}};

\begin{scope}[every path/.style={->}]
    \draw (p0) -- (p1);
    \draw (p1) -- (p0);
    \draw (p1) -- (p2);
    \draw (p2) -- (p1); 
    \draw (p2) -- (p3);
    \draw (p3) -- (p2); 
\end{scope} 

\end{tikzpicture}

\caption{Linkage between the risk neutral density, local volatility, flow, and drift.}
\label{threewaylink}

\end{figure}

\noindent
Given a discrete set of marginal distributions, i.e., partial information about $p(t,y)$, the Markov functional method leads to the construction of local volatility models of different characteristics, for example, step-wise time-homogeneous local volatility functions or those continuous in time and across marginal maturities. The resulting local volatility models can generate different implied skews for intermediate maturities. Instead of imposing a term structure of the flow function \textit{a priori}, selecting a drift function or a flow function in an optimal sense merits further study. An extension of the Markov functional approach beyond the local volatility models, e.g., path-dependent volatility models, is an interesting topic for future research.

\newpage

\begin{appendices}

\section{Fitting Implied Skews}
\label{skew_fitting}

\begin{figure}[!ht]
\centering
\includegraphics[scale=0.35]{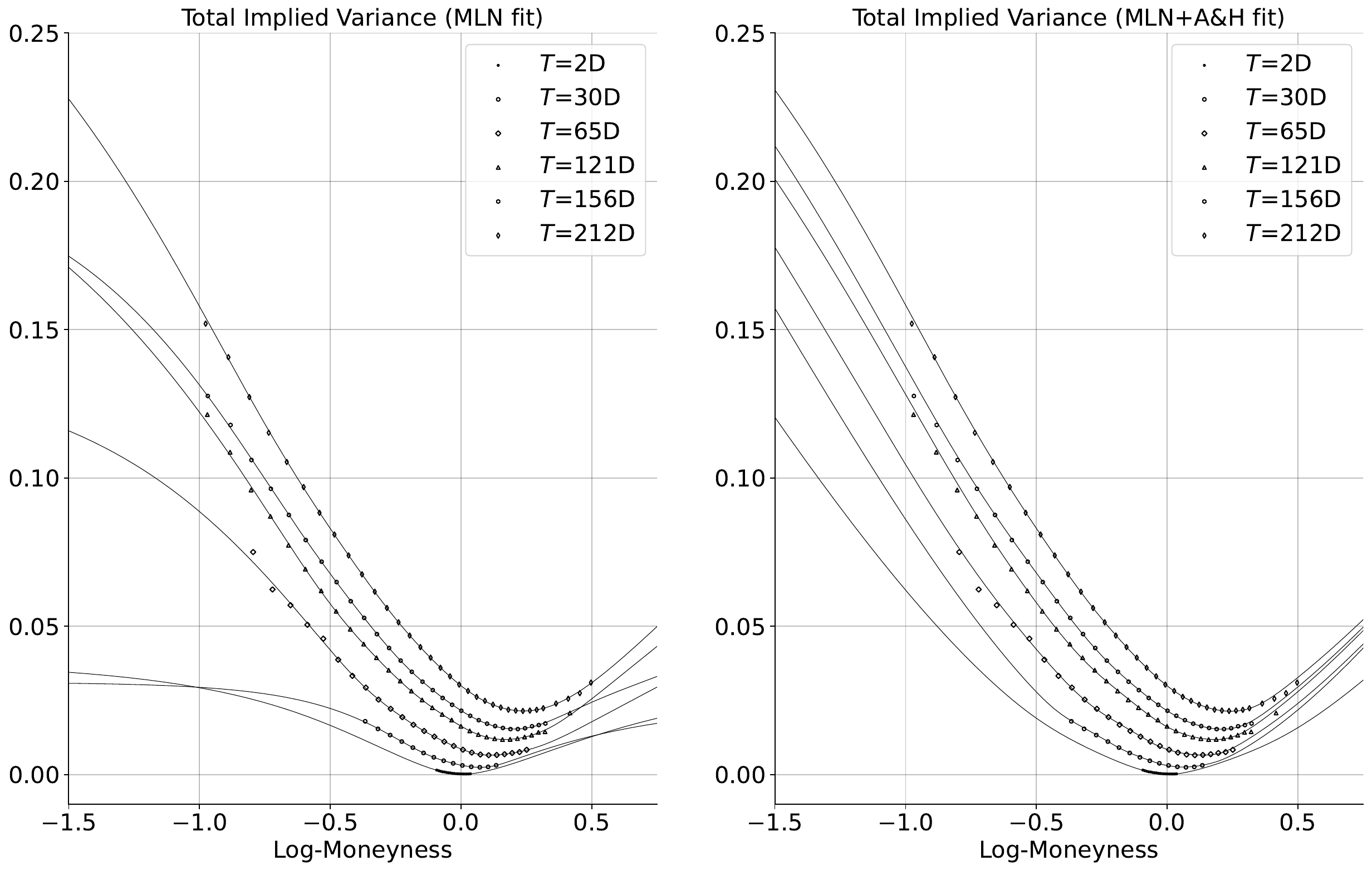}
\caption{Mixed lognormal fit to JPM options data on February 15, 2023. Left panel: fitting each slice of implied skews separately. Right panel: fitting the slices sequentially according to Algorithm \ref{mln_ah}.}
\label{jpm_fit}
\end{figure}

The volatility interpolation procedure developed in this paper depends on the availability of the marginal distribution functions $F_{\nu_i}(\cdot)$ to perform quantile-matching at the marginal maturities, as in Eq.(\ref{quantilematching}). To apply the interpolation procedure to market options data, an arbitrage-free (butterfly or calendar) fitting of the implied volatility skews is critical. Fitting implied skews slice by slice separately, however, suffers from calendar arbitrage, causing the subsequent volatility interpolation in the dimension of maturity to fail. This is shown in the left panel of Figure (\ref{jpm_fit}) for JPM options data on February 15, 2023, where one can see some crossings in the total implied variances of different maturities. One idea to remove the calendar arbitrage is to (1) use the Andreasen \& Huge (A\&H) local volatility \cite{andreasen2011} to enforce the convex order between successive marginal distributions; (2) perform an mixed lognormal (MLN) fit on the solution of the A\&H scheme. The procedure is described in Algorithm \ref{mln_ah}. The resulting fit of implied skews is shown in the right panel of Figure (\ref{jpm_fit}), where the crossings in total implied variance are removed at the expense of slightly reduced quality of fitting for some slices at large moneyness.


\begin{algorithm}[H]
\caption{Mixed Lognormal Fit of Andreasen \& Huge (A\&H) Solutions}\label{mln_ah}
\begin{algorithmic}
\State $\bullet$ Start from the skew at $T_0$: $C(T_0,K) = (S_0-K)^+$.
\State $\bullet$ For $i=0,1,2,\cdots$, assume the full skew $C(T_i,K)$ is available.
\State $\bullet$ Fit an MLN model $C^{\text{MLN}}(T_{i+1},K)$ to the prices $C(T_{i+1},K_j)$ at $T_{i+1}$.
\State $\bullet$ Compute A\&H local vol using $C(T_i,K)$ and $C^{\text{MLN}}(T_{i+1},K)$:
$$
\theta^2_i(K) \triangleq \frac{\frac{C^{\text{MLN}}(T_{i+1},K)-C(T_i,K)}{T_{i+1}-T_i}}{\frac{1}{2}\cdot \frac{\partial^2 C^{\text{MLN}}(T_{i+1},K)}{\partial K^2}}.
$$
\State $\bullet$ Solve the one-step implicit scheme of the forward PDE:
$$
\left[ 1 - \frac{1}{2}(T_{i+1}-T_i)\theta^2_i(K)\frac{\partial^2}{\partial K^2}  \right]\hat{C}(T_{i+1}, K) = C(T_{i}, K)
$$
\State $\bullet$ Optimize the MLN model parameters with respect to 
$$
\min_{\text{MLN parameters}} \sum_{j} | C(T_{i+1}, K_j) - \hat{C}(T_{i+1}, K_j) |^2.
$$
\State $\bullet$ Obtain the solution of the one-step forward PDE as $C(T_{i+1},K)$.
\State $\bullet$ Fit an MLN model to $C(T_{i+1},K)$ on the PDE grid.
\end{algorithmic}
\end{algorithm}

\end{appendices}

\end{document}